\documentclass[preprint,12pt]{elsarticle}

\usepackage{hyperref}
\usepackage{amstext}
\usepackage{graphicx}
\usepackage{epstopdf}
\usepackage{amssymb}
\usepackage{mathrsfs}
\usepackage{amsmath}
\usepackage{amsthm}
\usepackage{amsfonts}
\usepackage{verbatim}
\usepackage{color}
\usepackage{natbib}
\usepackage{stmaryrd}

\def\ang{\textup{\AA}}

\makeatletter
\newcommand{\myvec}[2][r]{%
  \gdef\@VORNE{1}
  \left(\hskip-\arraycolsep%
    \begin{array}{#1}\vekSp@lten{#2}\end{array}%
  \hskip-\arraycolsep\right)}

\def\vekSp@lten#1{\xvekSp@lten#1;vekL@stLine;}
\def\vekL@stLine{vekL@stLine}
\def\xvekSp@lten#1;{\def\temp{#1}%
  \ifx\temp\vekL@stLine
  \else
    \ifnum\@VORNE=1\gdef\@VORNE{0}
    \else\@arraycr\fi%
    #1%
    \expandafter\xvekSp@lten
  \fi}
\makeatother

\newtheorem{theorem}{Theorem}[section]

\newtheorem{algorithm}[theorem]{Algorithm}

\def\8{\infty}

\def\calE{\mathcal{E}}

\def\calP{\mathcal{P}}
\def\calR{\mathcal{R}}
\def\calT{\mathcal{T}}

\def\dsl{\llbracket}
\def\dsr{\rrbracket}

\def\ldsl{\left \llbracket}
\def\ldsr{\right \rrbracket}

\newcommand*{\dbl}{\{\mskip-5mu\{}
\newcommand*{\dbr}{\}\mskip-5mu\}}

\journal{Journal of Computational Physics}

\begin{document}

\begin{frontmatter}

\title{A Computational Model of Protein Induced Membrane Morphology with Geodesic Curvature Driven Protein-Membrane Interface}


\author{Y. C. Zhou\fnref{myfootnote}}
\address{Department of Mathematics, Colorado State University, Fort Collins, CO 80523}
\author{David Argudo, Frank Marcoline, Michael Grabe}
\address{Department of Pharmaceutical Chemistry and Cardiovascular Research Institute, University of California, San Francisco, 
CA 94143}
\fntext[myfootnote]{Corresponding author. Email: yzhou@math.colostate.edu}



\begin{abstract}
Continuum or hybrid modeling of bilayer membrane morphological dynamics induced by embedded proteins necessitates the identification
of protein-membrane interfaces and coupling of deformations of two surfaces. In this article we developed (i) a minimal total geodesic curvature 
model to describe these interfaces, and (ii) a numerical one-one mapping between two surface through a conformal mapping of each surface 
to the common middle annulus. Our work provides the first computational tractable approach for determining the interfaces between bilayer and 
embedded proteins. The one-one mapping allows a convenient coupling of the morphology of two surfaces. We integrated these two new developments 
into the energetic model of protein-membrane interactions, and developed the full set of numerical methods for the coupled system. Numerical
examples are presented to demonstrate (1) the efficiency and robustness of our methods in locating the curves with minimal total geodesic curvature
on highly complicated protein surfaces, (2) the usefulness of these interfaces as interior boundaries for membrane deformation, and (3)
the rich morphology of bilayer surfaces for different protein-membrane interfaces. 
\end{abstract}


\end{frontmatter}


\section{Introduction}
Lipid bilayer membranes are highly curved macromolecules whose curvatures have been long recognized as an essential structural
feature in key biological processes such as membrane trafficking, cytokinesis, infection, and cell motion. Biological
membranes, however, are far more complicated than simple bilayers because of the presence of non-lipid components, in particular, 
proteins. Measured by weight, the ratio of protein to lipid is about $0.2$ in myelin, while in the mitochondrial inner membrane
the ratio is about 3.0 \cite{GuidottiG1972a}. More recent measurements identified about $40$ different membrane proteins 
in the membrane of a synaptic vesicle with a diameter around $40$~nm \cite{TakamoriS2006a}. Generation, modulation, and 
maintenance of biological membrane curvature is therefore intrinsically coupled to the interactions between lipid bilayers 
and membrane proteins. These interactions also determine the conformational change of transmembrane segments of membrane 
proteins \cite{AndersenO2007a}, see for example the gating of mechanosensitive ion channels \cite{SukharevS2012a}. 
The importance and complexity of protein-mediated membrane morphological variation has thus fascinated investigators from various 
backgrounds.

In this article we develop a hybrid computational model of protein embedding into bilayer membrane for quantifying two critical 
geometric features of protein-membrane interactions. These are (i) the interface between bilayer 
and the embedded protein, and (ii) the surface morphology of the bilayer with the embedded protein. This interface sets a boundary 
for the lipid bilayer whose deformation will produce a tensional force on the boundary that shall displace the interface as a 
feedback, thus defining a bidirectional coupling between membrane morphological changes and the state of protein inclusion. 
Different treatments exist for these two features. Continuum \cite{LeeK2013a} or 
hybrid \cite{GrabeM2008a,ZhouY2010b,ChenX2010a,SigurdssonJ2013a,ArgudoD2017a} models usually treat the whole bilayer or 
its two individual leaflets as an elastic sheet using the classical Helfrich 
theory \cite{CanhamP1970,HelfrichW1973,EvansE1974} or Monge parameterization. An explicit specification of membrane edge and the
boundary condition for displacement are in these models. In fully atomistic \cite{KimT2012a} or coarse-grained \cite{TangY2006a,ChenX2010a} 
models it is not necessary to explicitly track the protein-membrane interface or membrane surface as they arise naturally as a result of trajectories of 
particles under simulations. However, it still remains a grand challenge for full atomistic or coarse-grained molecular dynamics simulators 
to model protein modulated membrane deformations or membrane mediated protein conformation changes of biologically relevant spatial and time scales 
\cite{FeigM2018a}. Continuum or hybrid models appear attractive for tackling these problems for they can be scaled to large domains 
and long time simulations at a relatively low computational cost. 

Description of protein membrane interfaces and boundary conditions for membrane elasticity vary in complexity and detail.
For example, in \cite{TangY2006a,BaviN2016a} the channel protein is represented as cylindrical rods located within (albeit without direction contact with) 
a pre-determined pore in an elastic sheet the modeling bilayer. The force between the protein rods and the membrane is described by Coulomb interactions, 
generating a bidirectional model for the gating of mechanosensitive ion channels. 
Consistent protein-membrane interface is found by using an immersed boundary method for proteins
presented as cylinders or cones \cite{SigurdssonJ2013a}. As a result, these models are able to generate rotational symmetrical membrane 
bending induced by proteins, but fail to reproduce realistic asymmetrical membrane deformations associated with anisotropic inclusions
of membrane proteins \cite{FournierJ1996a}. These asymmetrical deformations can be characterized by using the two distinct principal curvatures 
$\kappa_1 \ne \kappa_2$. These two curvatures are associated with the complex shape and specific orientation of membrane proteins, 
and are indispensable for the generation of negative Gaussian curvature required by all endocytosis and exocytosis processes of 
living cells \cite{NoguchiH2017a,ZhouY2019a}. Determination of protein-membrane interface, i.e., the contact curves between the protein and the two 
membrane surfaces, is at the center of the continuum or hybrid modelling of protein-membrane interactions \cite{CallenbergK2012a,SigurdssonJ2013a}. 

We model the protein-membrane contact lines as curves evolving on the 3-D protein surface driven by the geodesic curvature energy and the 
nonpolar energy. 
The geodesic curvature energy models the line tension at 
the contact curves between the protein and the membrane surfaces. This energy is locally minimized on membrane surface when the curve is a 
geodesic. 
To track the minimization of the geodesic curvature energy we adopt the approach in \cite{ZhouY2017a} to develop a surface phase field model where the evolution 
of the surface phase field function follows the gradient flow of the curvature energy. 
Furthermore, the bilayer has two high dielectric layers of polar headgroups that face the aqueous solution and a low dielectric hydrophobic core 
at the center. Inclusion of hydrophobic amino acids of the transmembrane proteins into the bilayer must go through the headgroup layers 
before getting into the energetically favorable hydrophobic core, and the final state of inclusion depends on the matching of hydrophobic domains
of both structures. The nonpolar energy, described by the scaled particle theory (SPT) \cite{PierottiR1976a,WagonerJ2006a}, is a function of the protein surface area exposed to 
the aqueous solvent, and thus can be represented using the surface phase field function as a force additional to the variation of the curvature energy functional.

The geodesic curvature flow in a Riemannian surface, also known as the curve shortening flow \cite{GraysonM1989a,GageM1990a}, is an important mathematical 
and computational tool in image processing, computer vision, and material sciences. The geodesic curvature flow equation, when posed in a level set formulation, 
is usually given by a highly nonlinear Hamilton-Jacobi equation for the level set function $\phi$:
\begin{equation} \label{eqn:geoflow_levelset}
\phi_t = | \nabla \phi | \nabla \cdot \left ( \frac{\nabla \phi}{\sqrt{| \nabla \phi |^2 + \beta}}  \right ),
\end{equation}
where $\beta>0$ is a small constant introduced to avoid division by zero \cite{OsherS1988a,SethianJ_FastMarchingBook,WuC2010a,LiuZ2017a}. The surface phase 
field model we developed previously for microdomain formation in bilayer membrane describes the similar evolution of the geodesic curvature flow and is numerically more 
tractable \cite{ZhouY2017a}. The limiting case of this surface phase field model at the vanishing intrinsic geodesic curvature of the microdomains 
is adopted to characterize the geodesic curvature flow in this study. 
The contact angle at the computed three-phase contact line (i.e., protein, solvent, and lipids) can be found 
using a generalized Young's formula \cite{WangY2001a,VazquezU2009a}, which prescribes the normal derivative of the displacement of the membrane surface. 
The position and the normal derivative comprise a full set of Dirichlet boundary condition on the protein-membrane interface. Along with the
far field boundary conditions (c.f (\ref{eqn:BC_far})) the fourth-order equation describing the membrane surface 
displacement \cite{GrabeM2008a,CallenbergK2012a} will be solvable. 

The displacements of the two membrane surfaces are not independent. Lipids are incompressible regardless of the gel or liquid phase in which they 
exist. This leads to the first constraint on the membrane displacement and has been characterized by the term penalizing the change of the membrane 
thickness \cite{ArgudoD2016a,ArgudoD2017a}. The inclusion of protein into the bilayer causes the lipid director, i.e., the average longitudinal axis
directed from the lipid head group to lipid tail, to deviate from their surface normal. This tilt deformation was 
recognized \cite{SafranS1986a,HammM2000a,JablinM2014a} but has not been considered in the continuum or hybrid modeling of bilayer membrane deformation 
until very recently \cite{RyhamR2016a}. Mismatch 
of the lipid directors of the two surfaces constitutes the second constraint for the coupling of the membrane displacement.
However, the inclusion of a transmembrane protein with arbitrarily complicated configuration into the bilayer makes it computationally challenging to establish
one-to-one correspondence between the two membrane surfaces. A previous practice decomposes the two membrane surfaces as matched and unmatched domains, while
the coupling of membrane displacement through the penalty of membrane thickness is only enforced in the matched domain \cite{ArgudoD2016a}. 
In this work we take advantage of the topological equivalence
of the two membrane surfaces with protein inclusion to map them onto the same annulus. This allows us to find the one-to-one correspondence between the
membrane surfaces, thus avoiding the splitting of matching and unmatched domains, and permitting the efficient enforcement of both constraints on the
membrane displacement. Finally, the inclusion of proteins with charged residues from the high dielectric solvent to the low dielectric bilayer core has an 
electrostatic barrier to pass through. The resolved protein-membrane interface and the membrane surfaces allow a convenient characterization of the 
dielectric interface and the corresponding electrostatic potential energy. 

The rest of the manuscript is organized as follows. In Section \ref{sect:model} a brief introduction of the transmembrane inclusion of protein
into lipid bilayer is followed by the energy functional formulation of solvated protein-membrane complex. The total energy consists of the
geodesic curvature energy, the coupled bilayer mechanical energy, nonpolar energy, and the electrostatic potential energy. We will
compute the weak derivatives of this functional with respect to the surface phase field function, the membrane displacements, and the electrostatic
potential. Corresponding partial differential equations (PDEs) will be derived. Numerical solutions of these coupled PDEs will be discussed in 
Section \ref{sect:numeric} where we will introduce the discontinuous Galerkin methods for solving the surface Cahn-Hilliard equation and the
membrane displacement. Mapping of the two membrane surfaces to the same annulus and the assembly of the surface mesh for 
the protein-membrane complex will also be presented. The proposed computational model is applied in Section \ref{sect:simulations} 
to the M2 proton channel, a protein in the influenza A virus envelope. 
The simulated protein-membrane interfaces and the membrane surface morphology will be examined against the experimental measurements. 
A summary and outline of future perspectives is presented in Section \ref{sect:summary}.

\section{Hybrid Model of the Solvated Bilayer Membrane with Transmembrane Protein} \label{sect:model}
\begin{figure}[!ht]
\begin{center}
\includegraphics[height=3cm]{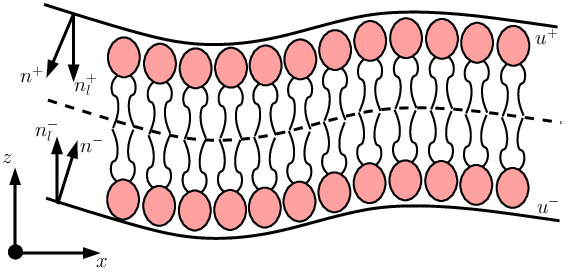}
\includegraphics[height=3cm]{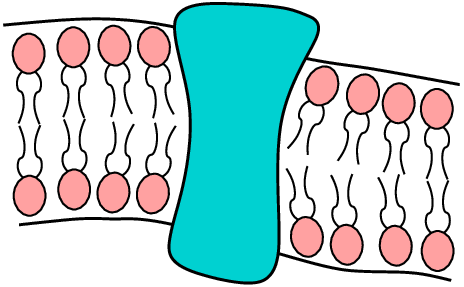}
\includegraphics[height=3cm]{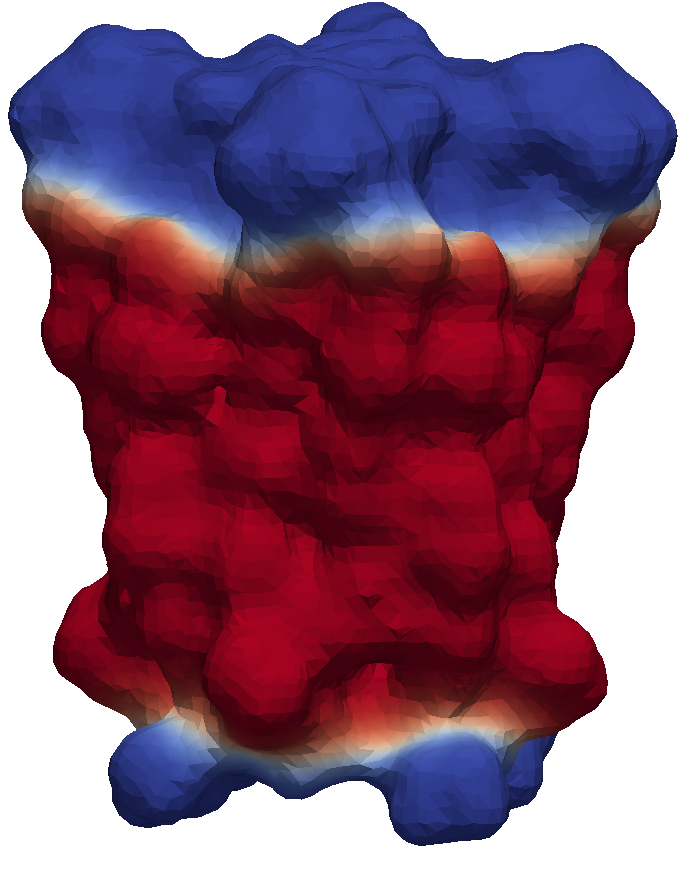}
\caption{Illustration of protein inclusion in a lipid bilayer. Left: the two membrane surfaces and the attached vectors. Middle: The protein inclusion causes
the unalignment of the lipid vectors from two bilayers near the inclusion. Right: The protein inclusion with two protein-membrane contact curves represented 
as the boundaries of a surface phase field function.}
\label{fig:complex_A}
\end{center}
\end{figure}

Consider a bilayer membrane as illustrated in Fig.~\ref{fig:complex_A} (left). The two surfaces, at height $u^+$ and $u^-$, 
represent the two leaflets of the bilayer stacked upon each other, with respective flat equilibrium height $\pm L_0$. The normals 
of the top ($+$) and bottom ($-$) surfaces, $n^+$ and $n^-$, may be not aligned with their respective top and bottom lipid directors, 
$n_l^+$ and $n_l^-$, when membrane surfaces are curved. Inclusion of the transmembrane protein causes the re-orientation of lipids near the protein, 
further deviating the lipid directors from their equilibrium orientation by tilt vectors $t^+$ and $t^-$ defined as 
\begin{equation} \label{eqn:nl}
t^{\pm} = n^{\pm} - n_l^{\pm}. 
\end{equation}
When the two lipid directors are not aligned as observed in case of protein inclusion, c.f. Fig.~\ref{fig:complex_A}(middle), the tilt energy 
could be significant. With the Monge parameterization of the surface heights $u^+,u^-$, the surface normals can be computed using
\begin{equation} \label{eqn:director}
n^{\pm} =(\mp \nabla u^{\pm}(x,y), \mp 1).  
\end{equation}
On the other hand, the transmembrane protein contacts with the two membrane surfaces at two closed curves, as shown in Fig.~\ref{fig:complex_A}(right), 
where they are modelled as zero-level sets of the phase field function $\psi$ defined on the protein surface, with red patch being the protein 
surface inside the bilayer while the blue patch in the aqueous solution exterior to the bilayer.

\subsection{Energetics of Solvated Protein-Membrane Complex}
The total free energy $G$ for the entire protein-membrane complex is defined as the summation of the electrostatic potential energy $G^e$, the
nonpolar energy $G^{np}$, the membrane elastic energy $G^{me}$, and the geodesic curvature energy $G^{geo}$ of protein-membrane interfaces:
\begin{equation} \label{eqn:total_G}
G = G^{e} + G^{np} + G^{me} + G^{geo}.
\end{equation}
Here the energy $G^{geo}$ is defined on the protein surface $\calP$ by 
\begin{equation} \label{eqn:Ggeo}
G^{geo} = \int_{\calP} K_l \left[ \xi \Delta_s \psi + \frac{1}{2 \xi} (1-\psi^2) \psi \right ]^2 ds
\end{equation}
where $\psi \in [-1,1]$ is the phase field function, $\Delta_s$ is the surface Laplacian, $\xi>0$ is a parameter adjusting the transition
of $\psi=-1$ in the bilayer to $\psi=1$ in the aqueous solution, and $K_l$ is the line tension coefficient. The square bracketed function 
is the approximation of the geodesic curvature using the surface phase field function \cite{ZhouY2017a}. Here it is associated with
line tension on the protein-membrane interface. 
The electrostatic potential energy $G^{e}$ is given by
\begin{equation} \label{eqn:Ge}
G^{e} = \int_{\Omega} \left[ \frac{\epsilon}{2} | \nabla \phi|^2 - q \phi - \kappa^2 \cosh(\phi) \right] dx,
\end{equation}
where $\phi$ is the electrostatic potential, $\Omega$ is a 3-D domain containing the membrane-protein complex, 
$\epsilon$ is the dielectric permittivity which takes different values in different molecular domains, $q$ is the spatial-dependent 
charge density, $\kappa$ is the ionic strength in the solvent (and thus is zero in proteins or lipid bilayers). 
Our model of electrostatics can be replaced by other variational formulations of implicit solvent models to include more sophisticated dielectric
effects such as nonlocal responses \cite{BardhanJ2011a,HildebrandtA2004a,XieD2013a}, finite particle size of ions and solvent \cite{LiB2009b,ZhouY2011a}, or 
surface charge or dipolar densities of bilayers \cite{ZhouY2010b,MikuckiM2014a}. The discontinuity of $\epsilon$ at the membrane surfaces 
shall induce dielectric surface forces, generating a driving force for membrane deformation.

The nonpolar energy $G^{np}$ is defined to be the work for transferring solute molecules between
polar and nonpolar solvents, given by a function of the Solvent Accessible Surface Area (SASA) 
\begin{equation} \label{eqn:Gnp}
G^{np} = \gamma_A (A_{m} - A_{s}), 
\end{equation}
where $\gamma_A$ is the constant characterizing the energy to transfer a unit molecular surface area from the polar to 
non-ploar solvent \cite{SitkoffD1996a}, and $A_{m}, A_{s}$ are SASAs 
in the interior and exterior of the bilayer, respectively. The term proportional to 
the molecular volume as seen in many nonpolar solvation energy models \cite{WagonerJ2006a,ChenZ2012a} 
is not considered here. The lateral pressure varies drastically across the bilayer because of 
the highly inhomogeneous molecular structure of the bilayer, and thus there does not exist a constant of proportionality similar 
to the uniform hydrostatic pressure in the aqueous solvent \cite{CantorR1997b,CantorR1999a,MarshD2013a}. Microscopic chemical potentials 
have been defined to quantify the dependence of lateral pressure on the cross-sectional area of the protein \cite{MarshD2007a,DrozdovaA2016a}. 
Energetically these efforts amount to relate the mechanical energy of the protein inclusion to the change of membrane morphology which will
be modelled by $G^{me}$. It is therefore necessary to remove the volume term of the nonpolar solvation energy in our model to avoid double counting.
With the surface phase field function $\psi$ we can conveniently approximate
\begin{equation} \label{eqn:SASA}
A_{m} = \int_{\calP} (\psi-1) ds, \quad A_s =-\int_{\calP} (\psi+1) ds. 
\end{equation}

The membrane elastic energy $G^{me}$ includes the energies associated with the fundamental modes of membrane deformations and lipid tilt:
\begin{align} \label{eqn:Gme}
G^{me} = & \int_{S} \frac{K_c}{2} \left[ (\nabla^2 u^+ + \nabla \cdot t^+ - J_0^+ )^2 + 
( \nabla^2 u^- - \nabla \cdot t^- - J_0^- )^2  \right] ds + \nonumber \\
         & \int_{S} \frac{K_G}{2} \left( K^- + K^+ \right ) ds + 
           \int_{S} \frac{\alpha}{2} \left( |\nabla u^-|^2 + |\nabla u^+|^2 \right ) ds + \nonumber \\
         & \int_{S} \frac{K_{\alpha}}{L_0^2} (u^+ - u^-)^2 ds + 
           \int_{S} \frac{K_t}{2} \left( |t^+|^2 + |t^-|^2 \right) ds + \nonumber \\
         & \int_{S} \frac{K_{tw}}{2} \left( |\nabla \times t^+|^2 + | \nabla \times t^-|^2 \right) ds + 
           \int_{S} \frac{K_e}{2} \left( |n^+ - n_0|^2 + |n^- - n_0|^2 \right) ds,
\end{align}
where the integrals in order represent the splay (mean curvature), saddle splay (Gaussian curvature), surface tension, compression of membrane surfaces and tilt-stretch, tilt-twist, 
and configurational entropic cost of lipid chains, with respective constant coefficients $K_c$ the bending modulus, $K_G$ the Gaussian modulus, $\alpha$ 
the surface tension coefficient, $K_{\alpha}$ the compression modulus, $K_t, K_{tw}, K_e$ the moduli of tilt, tilt-twist, 
and configurational confinement \cite{ArgudoD2016a}. $J_0^{\pm},n_0^{\pm}$ are the spontaneous
mean curvature and spontaneous director vector of the respective surfaces. These constants are assumed to be the same for two leaflets, though 
their dependence on the lipid composition can be introduced at the cost of computational complexity. $G^{\pm}$ are the Gaussian curvatures of 
the respective monolayers. The energetic modelling of membrane deformation modes has 
been classical and can be reduced to the Canham-Helfrich energy density \cite{CanhamP1970,HelfrichW1973,EvansE1974}. Lipid tilt has been recognized important 
recently \cite{HammM2000a,JablinM2014a} and proved to be critical in describing the energy pathway of membrane fusion during which lipid directors 
mismatch significantly \cite{RyhamR2016a}. For protein inclusions with arbitrary shape we expect the lipid directors would also differ significantly from
those away from the inclusion.

\subsection{Variational Principles and Nonlinear PDEs for the Hybrid Model}
Given the fixed 2-D domain $S$ for the Monge parameterization of the membrane surfaces, the static manifold $\calP$ modeling the protein surface, and the fixed 3-D domain $\Omega$ containing
the entire solvated protein-membrane complex, the total energy $G$ depends only on the membrane surface heights $u^{\pm}$ and the lipid directors $t^{\pm}$, 
noticing that the electrostatic 
potential $\phi$ depends on dielectric function $\epsilon$ which in turn depends on $u^{\pm}$. We shall compute the variational derivatives of $G$ with respect to $\phi,\psi$
and $u^{\pm},t^{\pm}$ to obtain the differential equations for $\phi, \psi$ and $u^{\pm}, t^{\pm}$. For $\phi$ we will have the 
Poisson-Boltzmann equation \cite{LiB2009a,WeiG2010a}
\begin{align} \label{eqn:PBE}
-\nabla \cdot (\epsilon \nabla \phi) + \kappa^2 \sinh(\phi) = q.
\end{align}
For $\psi$ we first compute 
\begin{equation}
g(\psi) = \frac{\delta G}{\delta \psi} = \frac{\delta G^{geo}}{\delta \psi} + \frac{\delta G^{np}}{\delta \psi} = 
\Delta_s W - \frac{1}{\xi^2} (3 \psi^2 -1 ) W + 2 \gamma_A,
\end{equation}
where
\begin{equation}
W = \xi \Delta_s \psi + \frac{1}{\xi} (1 - \psi^2).
\end{equation}
The evolution of the protein-membrane contact curves along the pseudo-time $t$ will follow the weak gradient flow of $\psi$:
\begin{equation} \label{eqn:ACE}
\frac{\partial \psi}{\partial  t} = -\frac{\delta G}{\delta \psi} = \Delta_s W_L + \frac{1}{\xi^2} W_L + \Delta_s W_N - \frac{3}{\xi^2} \psi^2 W + 
\frac{1}{\xi^2} W_N - 2 \gamma_A,
\end{equation}
where 
\begin{equation}
W_L = \xi \Delta_s \psi + \frac{1}{\xi} \psi, \quad W_N =-\frac{1}{\xi} \psi^3
\end{equation}
are respectively the linear and nonlinear components of $W$ such that $W=W_L + W_N$. This splitting will facilitate the numerical treatment 
of the Allan-Cahn equation (\ref{eqn:ACE}). Here the phase boundary is tracked without enforcing the total quantity of either phase, 
thus our model is free of the conservation constraints and the associated Lagrangian multipliers \cite{DuQ2011a,ZhouY2017a}. Our approach here is different from 
those based on the geodesic active contours \cite{CasellesV1997a,SpiraA2007a}, where an Euler-Lagrange equation is first derived for minimizing the 
curve energy in sharp interface formulation and then approximated using a level set form \cite{OsherS1988a,LiuZ2017a},

To derive the Euler-Lagrange equations for the unknowns $u^{\pm}, t^{\pm}$ we recognize that the dielectric function $\epsilon$ also depends on $u^{\pm}$, thus
we shall also consider the variational derivative of $G^{e}$ with respect to $u^{\pm}$. This derivative, by definition, is identical to the shape derivative
of the electrostatic potential energy under the smooth velocity field induced by the displacement $u^{\pm}$ \cite{MikuckiM2014a}, and thus gives the electrostatic 
forces $f_{e}$ on the membrane surfaces:
\begin{equation}
f^{\pm}_e = -\frac{\epsilon_s}{2} | \nabla \phi^{\pm}_s|^2 + \frac{\epsilon_m}{2} | \nabla \phi_m^{\pm} |^2 + 
\epsilon_m (\nabla \phi_s^{\pm} \cdot n^{\pm}) (\nabla \phi_m^{\pm} \cdot n^{\pm}) - \cosh(\phi),
\end{equation}
where $\phi_s, \phi_m$ are respectively the electrostatic potentials on the solution and 
membrane sides. The Euler-Lagrange equations for $u^-, t^-$ shall read
\begin{align}
K_c \Delta (\Delta u^- - \nabla \cdot t^- + J_0^-) - K_e \nabla \cdot (\nabla u^- - t^- + n_0^-) - 
\alpha \Delta u^- + 2 \frac{K_{\alpha}}{L_0^2} (u^- - u^+) & = f_e^-,  \label{eqn:u} \\
(K_t + K_e)t^- - K_e(\nabla u^- + n_0) + K_c \nabla (\Delta u^- - \nabla \cdot t^- + J_0) & = 0, \label{eqn:t}
\end{align}
noticing that $\epsilon$ does not depend on $t^-$ so the variational derivatives of $G^{e}$ do not generate a forcing term for
$t^-$. Equations for $u^+,t^+$ can be obtained similarly. The coupling of $u$ with $t$ in these equations motivates us to think of 
approaches to decouple them to facilitate numerical treatment. The assumption of infinite two-dimensional wall on
the surface of protein inclusion as put forward by May {\it et. al.} \cite{BohincK2003a,MayS2002a} is not applicable to our modeling
of the protein-membrane complex because it contradicts our consideration of the realistic protein geometry. Instead, we assume that
only within a lipid tail length the values of $K_e$ and $n_0$ are constants equaling to the values right at the protein-membrane 
interface in May's model \cite{MayS2002a}. Beyond that length different constants are taken, leading to the following piecewise 
definition of the two constants:
\begin{align}
& \frac{K_e}{k_BT} = \frac{12}{A_0}, \quad n_0 =-\frac{1}{2} n_p, & 0 \le r \le h_0, \\ 
& \frac{K_e}{k_BT} = \frac{3}{A_0}, \quad n_0 =0, & r > h_0, 
\end{align}
where $n_p$ is the unit binormal vector on the protein-membrane contact curves, $r$ is distance in the normal direction away from the 
protein surface, and $A_0$ is the cross-sectional area of each lipid chain. Consequently, one can remove $J_0^-,n_0^-$ from Equation (\ref{eqn:u}) and 
take divergence of Equation (\ref{eqn:t}) to get
\begin{align}
K_c \Delta (\Delta u^- - \nabla \cdot t^-) - K_e \nabla \cdot (\nabla u^- - t^- ) - 
\alpha \Delta u^- + 2 \frac{K_{\alpha}}{L_0^2} (u^- - u^+) & =-f_e^-,  \label{eqn:u_simp} \\
(K_t + K_e)\nabla \cdot t^- - K_e \Delta u^- + K_c \Delta (\Delta u^- - \nabla \cdot t^-) & = 0. \label{eqn:t_div}
\end{align}
We observe in Equations (\ref{eqn:u_simp},\ref{eqn:t_div}) that 
\begin{equation} \label{eqn:divt_a}
K_e \nabla \cdot (\nabla u^- - t^- ) + \alpha \Delta u^- - 2 \frac{K_{\alpha}}{L_0^2} (u^- - u^+) - f_e^-
= -(K_t + K_e)\nabla \cdot t^- + K_e \Delta u^-, 
\end{equation}
and thus we can solve for $\nabla \cdot t^-$:
\begin{equation} \label{eqn:div_t}
\nabla \cdot t^- = -\frac{\alpha}{K_t} \Delta u^- + \frac{2 K_{\alpha}}{K_t L_0^2} (u^- - u^+) + \frac{f_e^-}{K_t}.
\end{equation}
Put this back to Equation (\ref{eqn:u}) we will get
\begin{equation} \label{eqn:u-}
\Delta^2 u^- - \chi^- \Delta(u^- - u^+) - \gamma^- \Delta u^- + \beta^- (u^- - u^+) = \hat{f}_e^-,
\end{equation}
with
$$ \chi^- = \frac{2 K_{\alpha}}{L_0^2(K_t + \alpha)}, \quad \gamma^- = \frac{K_e}{K_c} + \frac{K_t \alpha}{K_c(K_t+\alpha)}, 
\quad \beta^- = \frac{2 K_{\alpha}}{L_0^2 K_c} \frac{K_t + K_e}{K_t + \alpha},$$
and 
$$ \hat{f}_e^- = \frac{K_c \Delta f_e^- - (K_t + K_e)f_e^-}{K_c(K_t + \alpha)}.
$$
A similar equation for $u^+$ shall read
\begin{equation} \label{eqn:u+}
\Delta^2 u^+ - \chi^+ \Delta(u^- - u^+) - \gamma^+ \Delta u^+ + \beta^+ (u^- - u^+) = \hat{f}_e^+,
\end{equation}
with
$$ \chi^+ =\frac{-2 K_{\alpha}}{L_0^2(K_t - \alpha)}, \quad \gamma^+ = \frac{K_e}{K_c} + \frac{K_t \alpha}{K_c(K_t-\alpha)}, 
\quad \beta^+ = \frac{2 K_{\alpha}}{L_0^2 K_c} \frac{K_t - K_e}{K_t - \alpha},$$
and
$$
\hat{f}_e^+ =-\frac{K_c \Delta f_e^+ - (K_t - K_e)f_e^+}{K_c(K_t - \alpha)}.
$$

Proper boundary conditions are needed for the solutions of equations we derived above. Electrostatic potential induced by fixed charges of the protein 
in the aqueous solution where $\epsilon_w \approx 80$ can be used as the approximate boundary conditions of the 
Poisson-Boltzmann equation (\ref{eqn:PBE}) on $\partial \Omega$ \cite{ZhouY2008a,LuB2007e}. The surface Allan-Cahn equation does not need
a boundary condition as it is defined on the closed protein surface $\calP$. For the coupled fourth-order equations (\ref{eqn:u-},\ref{eqn:u+}) 
we adopt the fixed boundary condition (homogeneous Dirichlet boundary condition for biharmonic equations) on the boundary 
far away from the protein inclusion:
\begin{equation} \label{eqn:BC_far}
u^{\pm} = 0, \quad \nabla u^{\pm} \cdot n = 0,
\end{equation}
indicating that the membrane is free of deformation on $\partial \Omega$. 
The protein-membrane interfaces are indeed the contact curves of three phases: the aqueous
solvent, the protein, and the bilayer. Wang {\it et. al.} \cite{WangY2001a} proposed a modification of 
the classical Young's contact angle \cite{VazquezU2009a} to include the tension energy of the contact line among solid-liquid-vapor phases:
\begin{equation}
\gamma_{lv} \cos \theta = \gamma_{sl} - \gamma_{sv} + \frac{\tau}{R}
\end{equation}
where $\theta$ is the  $\gamma_{lv},\gamma_{sl},\gamma_{sv}$ are respectively the surface tensions of liquid-vapor, solid-liquid, and solid-vapor interfaces, 
$\tau$ is the line tension and $R$ is the radius of the base of the liquid drop in contact with the solid. It is interesting to recognize that
if the base contact curve is a geodesic then $1/R$ shall be zero and the classical Young's contact angle is reproduced. Therefore at our
protein-water-lipid interface we will have
\begin{equation}
\cos \theta =  \frac{\gamma_{pl}}{\gamma_{lw}} - \frac{\gamma_{pw}}{\gamma_{lw}}.
\end{equation} 
The protein-water interface tension was well-estimated \cite{DerA2007a}, and given the fluidity of the lipid molecules in the two-dimensional surface we 
postulate that $\gamma_{pl} \approx \gamma_{pw}$ thus $\cos \theta = 0$, leading to an estimate that $\theta = \pi/2$. This suggests that the two membrane 
surfaces are locally orthogonal to the protein surface at the protein-membrane contact curves, thus the following fixed boundary conditions will be adopted
\begin{equation} \label{eqn:BC_near}
u^{\pm} = u_{B}^{\pm}, \quad \nabla u^{\pm} \cdot n = \theta_{p},
\end{equation}
where $u_{B}^{\pm}$ are the positions of the protein-membrane contact curves and $\theta_p$ is the projection of the unit normal direction at the contact
curves onto the common $x-y$ base plane on which the Monge parameterizations of the two membrane surfaces are defined.

\section{Numerical Techniques for the Nonlinear PDEs} \label{sect:numeric}
In this section we discuss the numerical solutions of the differential equations and the convergent iterations for their coupling. The evolving 
protein-membrane contact curves on the protein surfaces present varying dielectric interface for the Poisson-Boltzmann equation, so a numerical 
method that does not require the regeneration of interface conforming 3-D mesh is preferred. Here we adopt a tree-code accelerated boundary 
integral method for solving the Poisson-Boltzmann equation \cite{GengW2013a,ChenJ2018a}. The molecular surface mesh of the protein-membrane 
complex will be updated when the protein-membrane contact curves evolve or the membrane surfaces are deformed with the given contact curves.

Our numerical treatments of the problem are focused on the solutions of the surface Allan-Cahn equation (\ref{eqn:ACE}), the surface 
deformation equations (\ref{eqn:u-},\ref{eqn:u+}), and their convergent coupling with the Poisson-Boltzmann equation. The $C^0$ interior penalty 
discontinuous Galerkin (DG) method \cite{BrennerS2012a,ZhouY2017a} is not applicable to the surface deformation equations as they only admit 
Cahn-Hilliard type boundary conditions while we have the Dirichlet boundary conditions here. Based on the DG method in \cite{GeorgoulisE2008a} we 
develop an alternative $C^0$ interior penalty method to solve the deformation equations (\ref{eqn:u-},\ref{eqn:u+}). Consider the following prototype of the 
surface deformation equation with Dirichlet boundary conditions $(g_D,g_N)$:
\begin{equation} \label{eqn:u_model}
\Delta^2 u + \beta_1 \Delta u + \beta_2 u = f, \quad u = g_D,  ~ \nabla u \cdot n = g_N,
\end{equation}
on the annulus domain $\calR$, c.f. Fig.~\ref{fig:2atm}(right). Let $\calT_h$ be a regular simplifical triangulation of $\calR$. We will use
the following standard notions for defining discontinuous Galerkin methods:
\begin{itemize}
\item $h_{T}$: diameter of the triangle $T \in \calT_h$, where $h = \max_{T \in \calT_h} h_T$,
\item $\calE_h$: set of edges of the triangles in $\calT$,
\item $\calE_h^i$: subset of $\calE_h$ consisting of edges interior to $\calR$,
\item $\calE_h^b$: subset of $\calE_h$ consisting of edges on the boundary $\partial \calR$,
\item $|T|,|e|$: the area of the triangle $T$, and the length of the edge $e$.
\end{itemize}
On an interior edge $e$ that is shared by the two triangles $T_{\pm}$ we define the normal vector $n_e$ pointing from $T_-$ to $T_+$ 
and the jump and average quantities:
$$ \dsl v \dsr = u_+ - u_-, \qquad \dbl u \dbr = \frac{1}{2} (u_+ + u_-)$$
for $u \in H^1(\calR,\calT_h)$. Here the subscripts $\pm$ signify the limit values of the function at the common edge $e$ on the 
triangles $T_{\pm}$. These definitions are extended to the boundary edge $e \in \calE^b_h$ where
$$ \dsl v \dsr = \dbl u \dbr = u|_e.$$
We associate with $\calT_h$ a standard quadratic Lagrange finite element space $V_h = \{ v \in C(\bar{\calR}): v_T = v|_T \in P_2(T), 
\forall T \in \calT_h \}$. The approximate solution $u_h \in V_h$ to Equation (\ref{eqn:u_model}) is given by 
\begin{equation} \label{eqn:DG}
\mathcal{B}(u_h,v_h) = l(v_h), \quad \forall ~ v_h \in V_h,
\end{equation} 
where the bilinear form $\mathcal{B}(\cdot,\cdot)$ and the linear functional $l(\cdot)$ are given, respectively, by
\begin{align}
\mathcal{B}(u_h,v_h) =  &  \sum_{T \in \calT_h} \int_{T} \big ( \Delta u_h \Delta v_h + \beta_1 \nabla u_h \cdot \nabla v_h + \beta_2 u_h v_h \big ) dx +  \nonumber \\
                      + & \sum_{e \in \calE^b_h} \int_{e} \alpha_1 u_h v_h ds  \nonumber \\
                      + & \sum_{e \in \calE_h} \int_{e} \left ( -\ldsl \frac{\partial u_h}{\partial n_e} \ldsr \dbl \Delta v_h \dbr - 
\ldsl \frac{\partial v}{\partial n_e}  \ldsr + 
\alpha_2 \ldsl \frac{\partial u_h}{\partial n_e} \ldsr \ldsl \frac{\partial v_h}{\partial n_e} \ldsr \right ) ds,  \label{eqn:bilinear} \\
l(v_h) = & \sum_{T \in \calT_h} \int_{T} f v_hdx + 
\sum_{e \in \calE_h^b} \int_{e} \left ( -g_N \Delta v_h + \alpha_1 g_D v_h + \beta_1 g_N \frac{\partial v_h}{\partial n_e} \right ) ds. \label{eqn:linear}
\end{align}
Here the penalty parameters $\alpha_1, \alpha_2$ are given by
\begin{equation} \label{eqn:penalty_const}
\alpha_1 = C_1 h^{-3}, \quad \alpha_2 = C_2 h^{-1}
\end{equation}
for sufficiently large positive constants $C_1,C_2$ that depend only on the triangulation $\calT_h$. The optimal convergence of 
this interior penalty DG method and the generalization of the analysis to higher order basis functions are given in \cite{GeorgoulisE2008a}.

Coupling of the deformation of two membrane surfaces demands the two equations (\ref{eqn:u-},\ref{eqn:u+}) to be defined on the same domain. However, 
this appears not the case because of the different contact curves of the transmembrane protein with two membrane surfaces. Since the two surfaces 
are expected to match point-wisely we are motivated to define smooth mapping between the middle plane and the base planes of two membrane surfaces. 
We define the interior boundary of the middle plane as a circle whose radius and height are the respective average radius and height of the two contact curves. 
The exterior boundary is at the middle of the exterior boundaries of the two surfaces. These mappings are numerically established by solving 
two Winslow equations \cite{Girdgeneration_KS} on the annulus domain $\calR$ in the middle plane. The meshes are locally refined toward the
contact curves where larger membrane deformations are expected, c.f. Fig.~\ref{fig:2atm}(right). 

Approximation of the surface biharmonic and surface Laplacian for the solution of Equation (\ref{eqn:ACE}) can be done 
as in (\ref{eqn:bilinear},\ref{eqn:linear}), without terms related to the boundary conditions though, as the evolution of phase field function
is on a closed 2-manifold $\calP$ approximated by a simplicial surface triangulation. There are many energy stable time discretization schemes
developed for the phase-field models, including convex splitting methods \cite{ElliottC1993a,ShenJ2012a,HuZ2009a}, 
exponential time discretization schemes \cite{WangX2016a}, IEQ schemes \cite{YangX2017a}, and semi-implicit methods \cite{ChenR2015a}. Many of these
methods introduce a chemical potential  as an intermediate variable to reduce the fourth-order Allan-Cahn or Cahn-Hilliard equation to 
two coupled second-order equations, which is not consistent with our discontinuous Galerkin approximation. Here the implicit method 
we developed for the surface Allan-Cahn equation \cite{ZhouY2017a} will be adapted. The method begins with a Crank-Nicolson approximation of 
the time derivation in Equation (\ref{eqn:ACE}), giving rise to
\begin{equation} \label{eqn:psi_CN}
\frac{\psi_{n+1} - \psi_n}{\Delta t} + g(\psi_{n+1}, \psi_n) + 2 \gamma_A = 0,
\end{equation} 
where $\Delta t$ is the time increment. The average function $g(\psi_{n+1},\psi_n)$ is defined as
\begin{align}
g(\psi_{n+1},\psi_{n})  = &\frac{1}{2} \Delta_s \left( f_c(\psi_{n+1}) + f_c(\psi_n) \right) \nonumber \\ 
- & \frac{1}{2\xi^2} \left( \psi^2_{n+1} + \psi_{n+1} \psi_n + \psi_n^2 -1) \big( f_c(\psi_{n+1}) + f_c(\psi_n) \big) \right)
\end{align}
with
$$ f_c(\psi) = \xi \Delta_s \psi + \frac{1}{\xi^2}(1 - \psi^2).$$
To solve Equation (\ref{eqn:psi_CN}) which is nonlinear and implicit in $\psi_{n+1}$, we define inner iterations on the variable $\Psi_{m}$
which is expected to converge to $\psi_{n+1}$ as $m \rightarrow \infty$. Replacing all linear and nonlinear terms of $\psi_{n+1}$ respectively 
using $\Psi_{m+1}$ and $\Psi_m$, we can get the following inner iteration
\begin{equation} \label{eqn:Psi}
\frac{\Psi_{m+1} - \psi_n}{\Delta t} + g(\psi_n,\Psi_m,\Psi_{m+1}) + 2 \gamma_A = 0,
\end{equation}
with the new average function
\begin{align}
g(\psi_{n},\psi_n,\Psi_{m+1}) = & \frac{1}{2} \Delta_s \tilde{f}_c(\psi_n,\Psi_m,\Psi_{m+1}) - 
\frac{1}{2 \xi^2} \left( \Psi_m^2 + \Psi_m \psi_n + \psi_n^2 - 1) (f(\Psi_m) + f(\psi_n) \right)
\end{align}
where
$$\tilde{f}_c(\psi_n,\Psi_m,\Psi_{m+1}) = \frac{\xi}{2} \Delta_s (\Psi_{m+1} + \psi_n) - \frac{1}{4 \xi} (\Psi_m + \psi_n^2 -2)(\Psi_m + \psi_n).$$ 
Now the equation (\ref{eqn:Psi}) is linear in $\Psi_{m+1}$ as we expect: 
\begin{align}
\left( 1 + \frac{\xi \Delta t}{2} \Delta_s^2 \right) \Psi_{m+1} = & \psi_n -\frac{\xi \Delta t}{2} \Delta_s^2 \psi_n 
+ \frac{\Delta t}{4 \xi} \Delta_s (\Psi_m^2 + \psi_n^2 -2)(\Psi_m + \psi_n)  \nonumber \\
+ & \frac{\Delta t}{2 \xi^2} \Delta_s (\Psi_m^2 + \Psi_m \psi_n + \psi_n^2)\left( f_c(\Psi_m) + f_c(\psi_n) \right) + 2 \Delta t \gamma_A.  \label{eqn:ACE_final}
\end{align}

Since the surface phase field model can only find the local minimum of the geodesic curvature energy, we will scan a range of initial 
protein-membrane contact curves to finally find the global minimum of the total energy $G$ defined in Equation (\ref{eqn:total_G}).
The full algorithm of our energetic model is summarized below:
\begin{algorithm} \label{alg:full_alg}
\begin{enumerate}
\item Define a proper range for the initial position of bilayer with respect to the given membrane protein. 
Divide the range into small intervals for the loop in Step 2 below.
\item For each initial position of the bilayer:
\begin{enumerate}[(i)]
\item Update the protein-membrane interfaces by solving the surface phase field equation (\ref{eqn:ACE_final}).
\item Update the mappings from the middle plane to the base planes of two membrane surfaces.
\item Update the electrostatic surface forces by solving the Poisson-Boltzmann equation with updated membrane surfaces.
\item Update the membrane surfaces by solving the coupled surface deformation equations.
\item Return to Step (ii) until convergence. Save total energy $G$.
\end{enumerate}
\item Choose the minimum $G$ and the corresponding geometry of protein-membrane complex.
\end{enumerate} 
\end{algorithm}

\section{Model Validation and Computational Simulations} \label{sect:simulations}
In this section we shall first validate our geodesic curvature modelling of the protein-membrane interfaces.
We will then present examples of mapping the base planes of membrane surfaces to the middle annulus.
We will finally apply these validated modules of our energetic model for the determination of inclusion state of the transmembrane protein.

\subsection{Validation of geodesic curvature model}
\begin{figure}[!ht]
\begin{center}
\includegraphics[width=2.5cm]{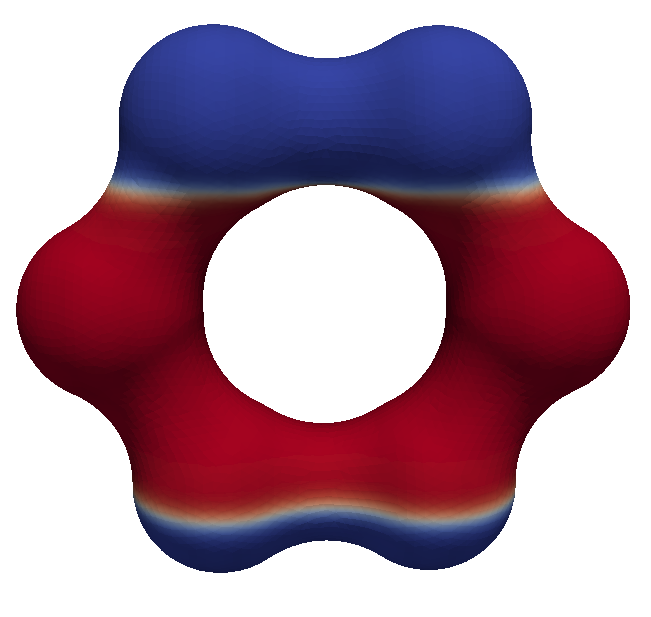}
\includegraphics[width=2.5cm]{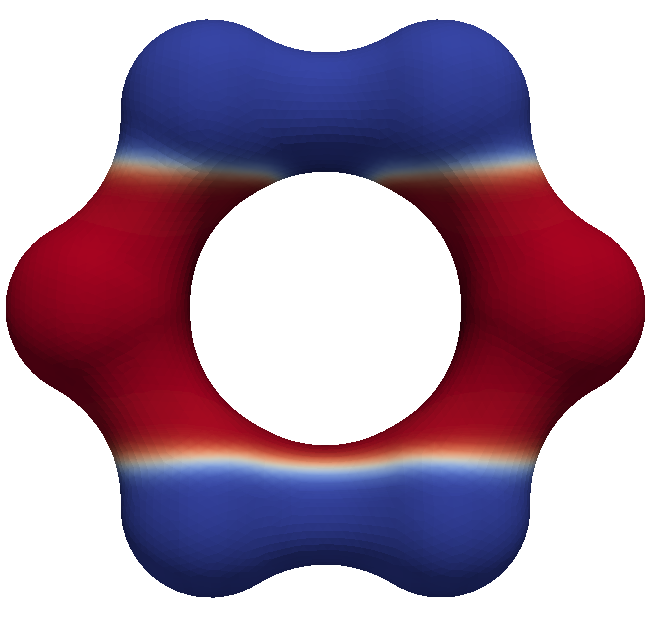}
\includegraphics[width=2.5cm]{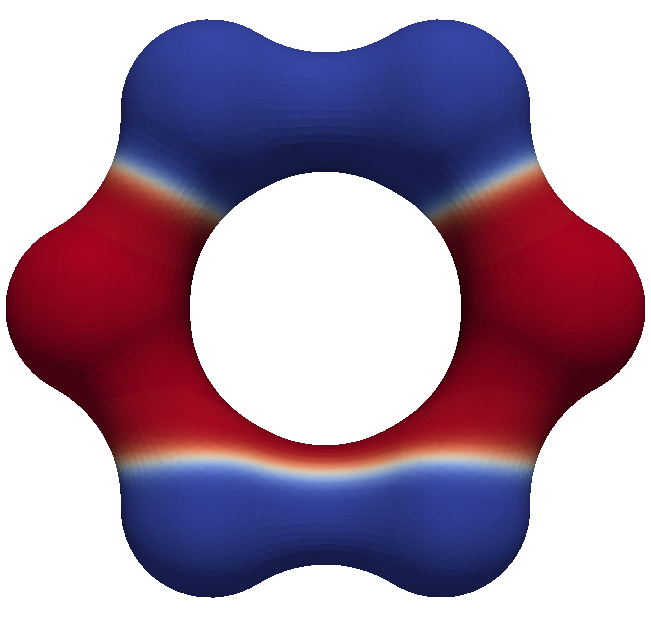}
\includegraphics[width=2.5cm]{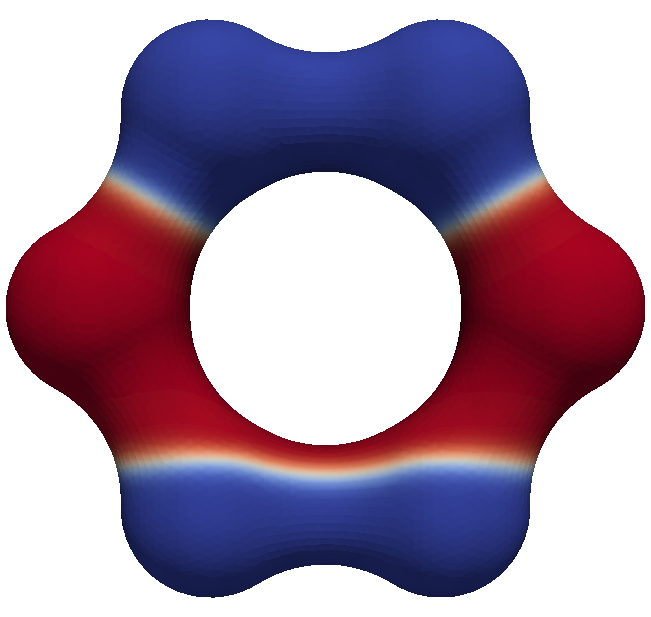}
\includegraphics[width=2.5cm]{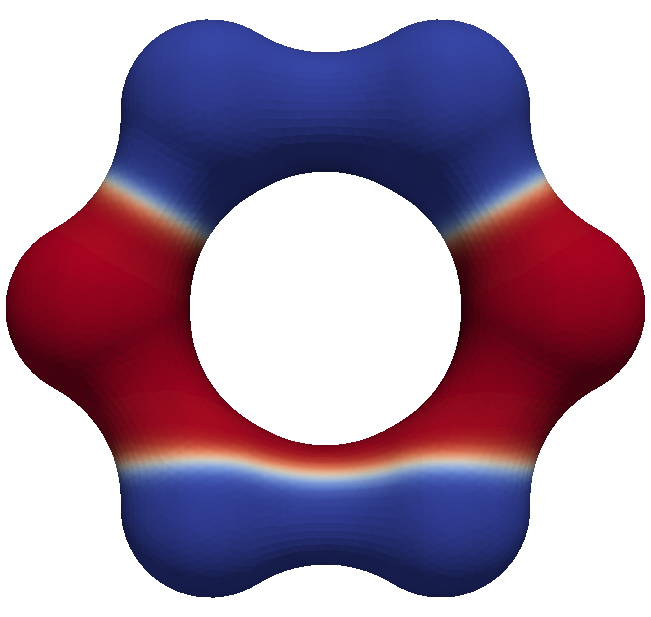} \\
\includegraphics[width=2.5cm]{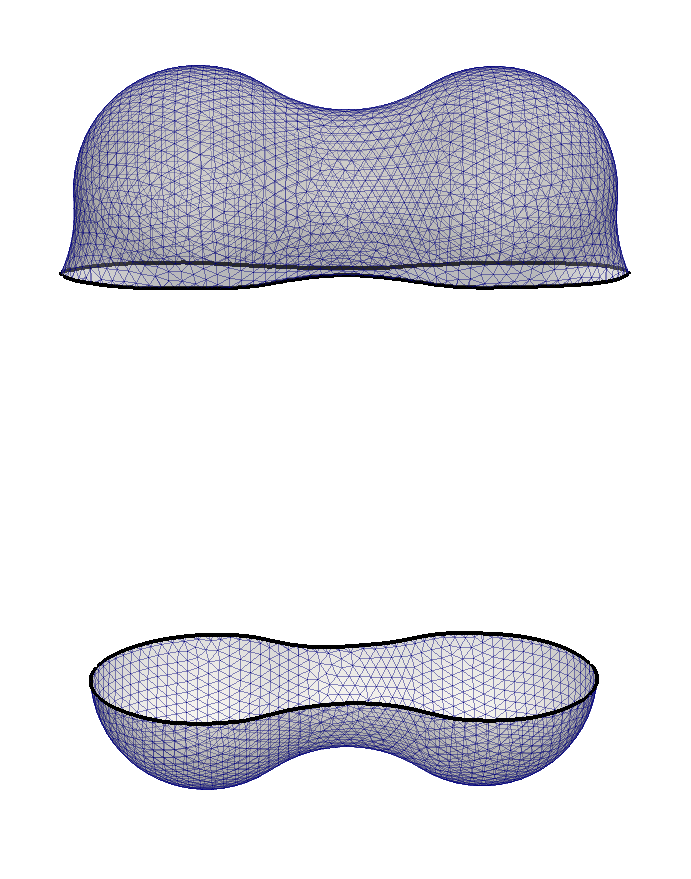}  
\includegraphics[width=2.5cm]{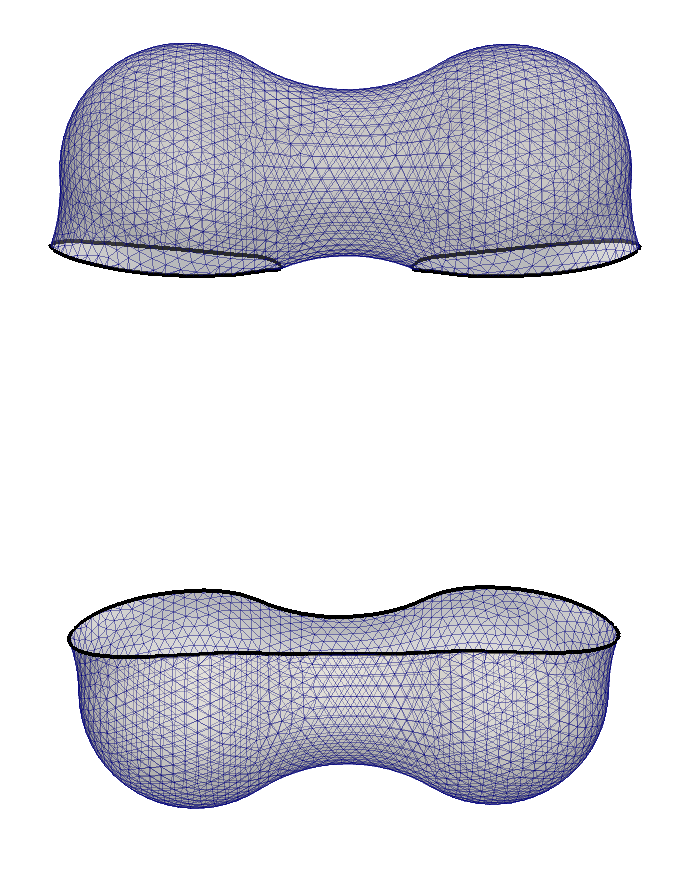} 
\includegraphics[width=2.5cm]{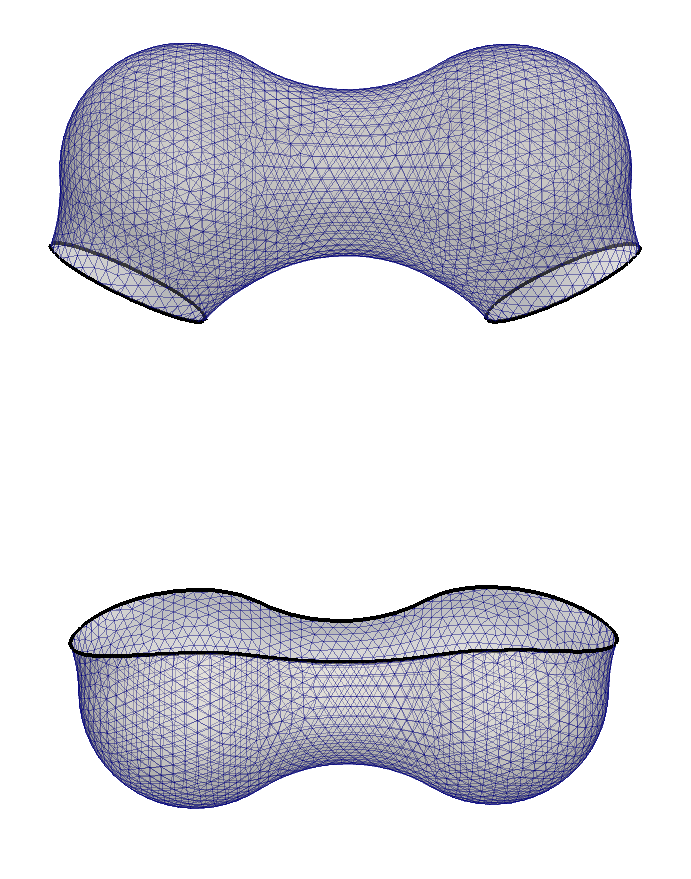} 
\includegraphics[width=2.5cm]{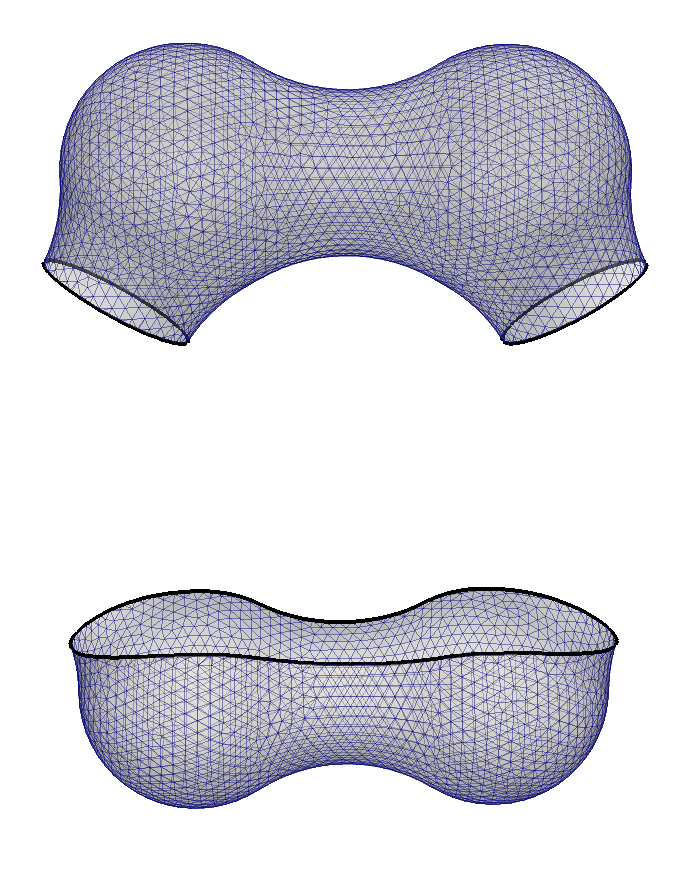} 
\includegraphics[width=2.5cm]{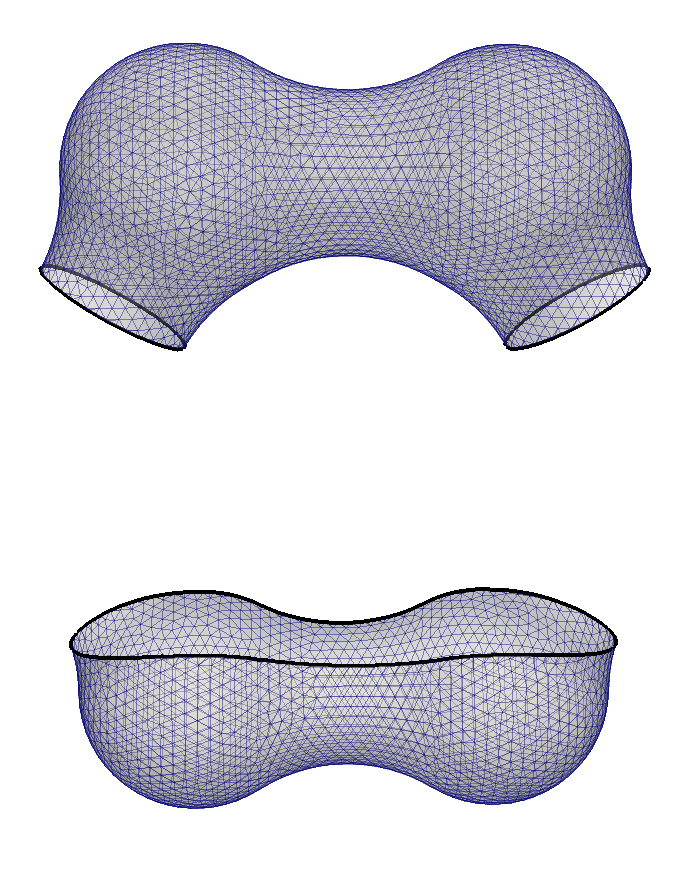} 
\caption{Snapshots in the surface phase boundary evolution on the model benzene ring at time $t=0,0.1,5,12,18$. 
Surface triangular mesh is quasi-uniform with $8124$ vertices
and $16248$ triangles. Time increment $\Delta t = 0.01$, $\xi=0.1$. Top: Plots of surface phase field function. Bottom: Surface mesh for the 
phase $\psi=-1$ highlights the phase boundaries with different history of topological change during the phase boundary evolution.}
\label{fig:ring_case1}
\end{center}
\end{figure}

We first apply our geodesic curvature model to capture the curve on a molecular surface with the well-identified local minimal geodesic curvature. 
The molecule mimics the carbon backbone of a benzene ring with six atoms in $y$-plane of unit radius respectively centered at
$(x,z)=(2.5,0)$, $(1.25,2.165)$,\\
$(-1.25,2.165)$, $(-2.5,0)$, $(-1.25,-2.165)$ and $(1.25,-2.165)$. In the first test, shown in 
Fig.~\ref{fig:ring_case1}, we chose the initial phase field $\phi=1$ 
for $z_{ini} \in [-1.62,1.52]$ and $\phi=-1$ elsewhere. This asymmetrical initial condition allows the two initial phase boundaries to evolve toward local 
minima of different topology. The upper phase boundary evolves downward, quickly splitting into two closed curves each of which evolves toward 
its own local minimum of geodesic curvature energy. The lower phase boundary below, in contrast, is able to maintain its topology and evolves upward 
gradually settle on a local minimum. The initial position $z_{ini}=1.52$ for the upper phase boundary is near the critical point. An initial phase boundary 
above this critical $z$-position will not split during the evolution but converges to a final position symmetrical to the convergent position of the phase 
boundary initialized at $z_{ini}=-1.62$ below.

\begin{figure}[!ht]
\begin{center}
\includegraphics[width=2.5cm]{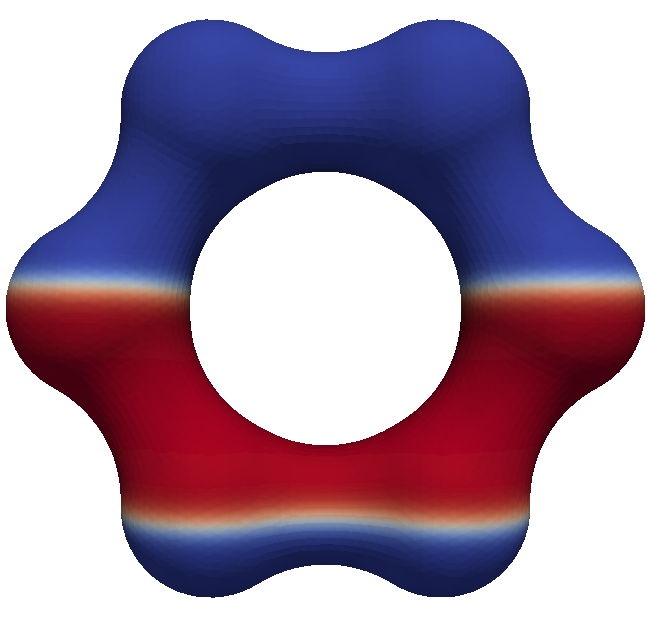} 
\includegraphics[width=2.5cm]{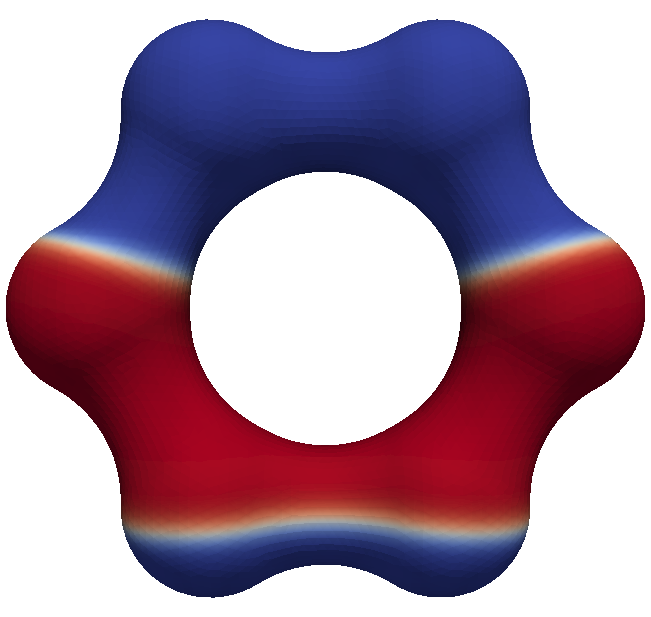} 
\includegraphics[width=2.5cm]{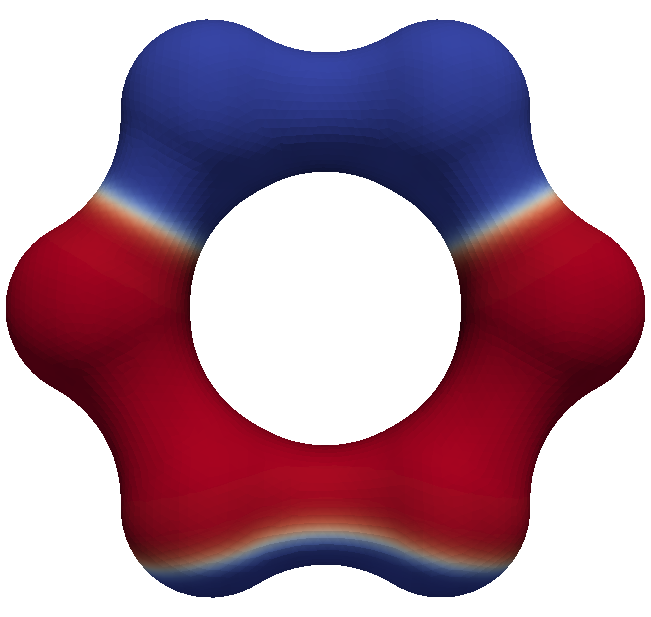}
\includegraphics[width=2.5cm]{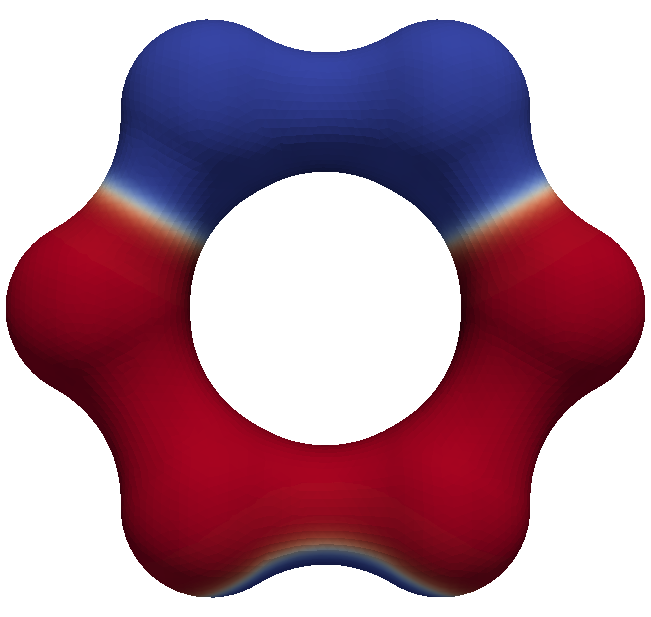}
\includegraphics[width=2.5cm]{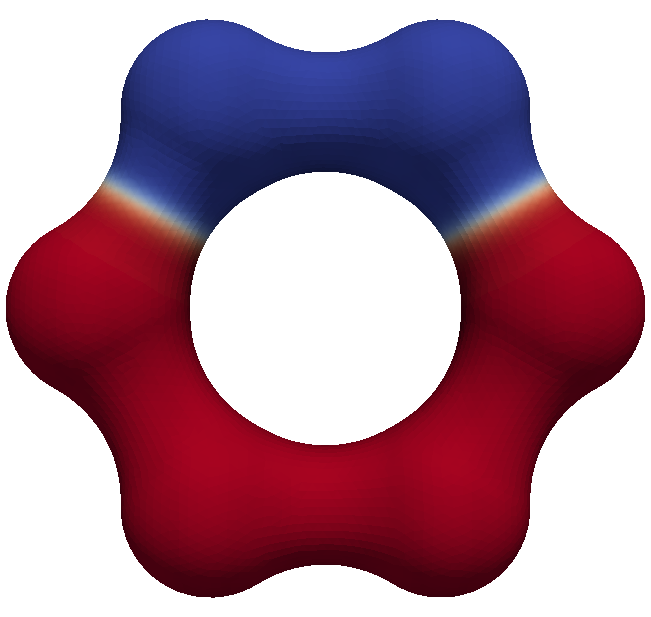} \\
\includegraphics[width=2.5cm]{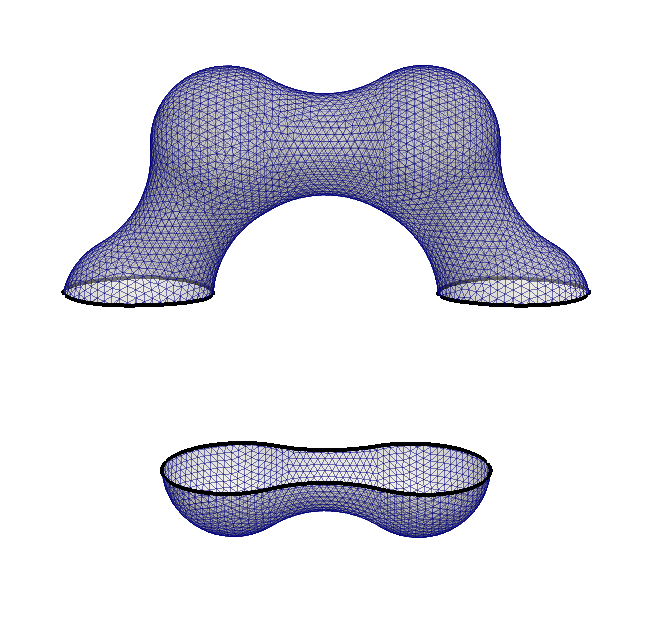}  
\includegraphics[width=2.5cm]{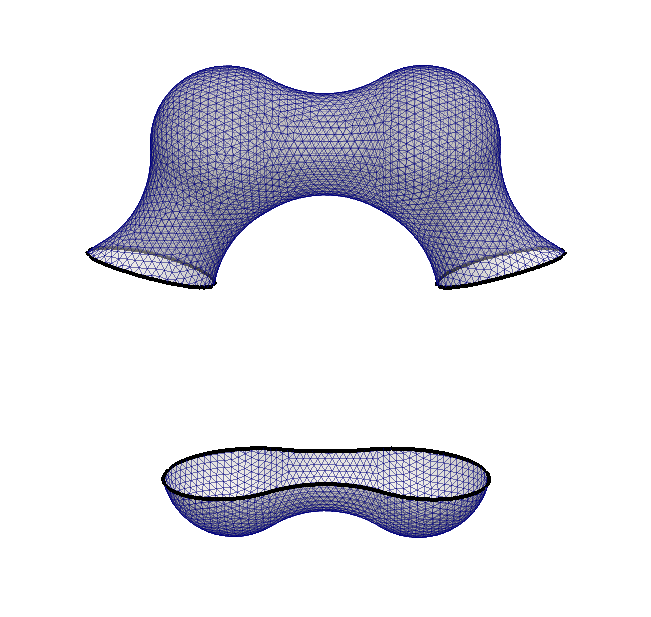}  
\includegraphics[width=2.5cm]{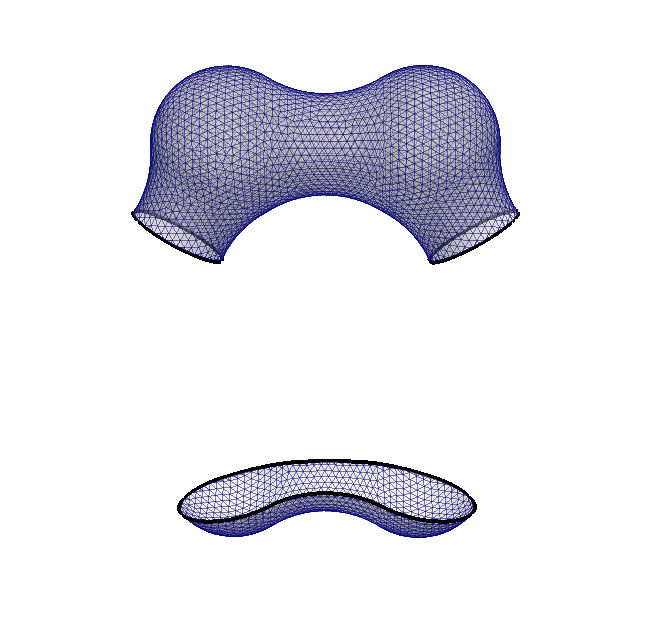}  
\includegraphics[width=2.5cm]{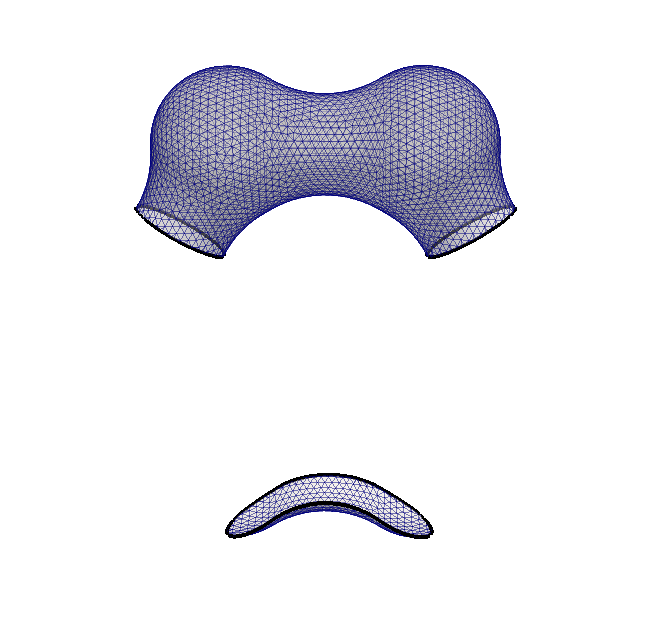}  
\includegraphics[width=2.5cm]{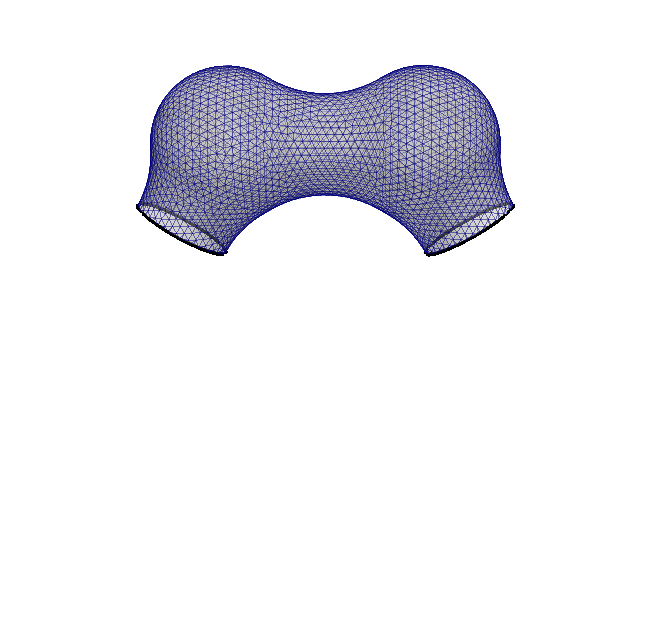}  
\caption{Snapshots in the surface phase boundary evolution on the model benzene ring at time $t=0,6,14,18,26$. 
Mesh and parameters are identical to the case in Fig.~\ref{fig:ring_case1}. 
Top: Plots of surface phase field function. Bottom: Surface mesh for the phase $\psi=-1$ highlights the phase boundaries with 
different history of topological change during the phase boundary evolution. The lower phase boundary disappears 
finally as a result of geodesic curvature minimization.}
\label{fig:ring_case2}
\end{center}
\end{figure}

In the second test, shown in Fig.~\ref{fig:ring_case2}, the surface phase field is initialized with phase boundaries at $z_{ini}=-2.22$ and $0.2$, 
respectively. The upper phase boundary evolves upward and converges to the same local minimum found by the upper phase boundary in 
Fig.~\ref{fig:ring_case1}. The lower phase boundary, which is initially $0.6$ below the lower phase boundary in the first test, evolves downward, 
shrinks and finally disappears, achieving a local minimum geodesic curvature energy of zero. 

\begin{figure}[!ht]
\begin{center}
\includegraphics[height=3.5cm]{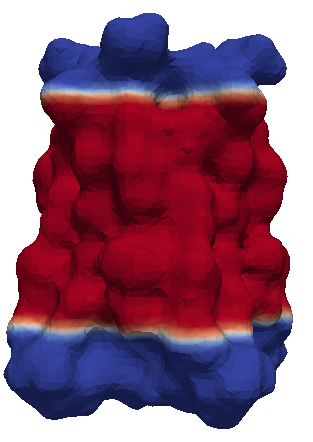} 
\includegraphics[height=3.5cm]{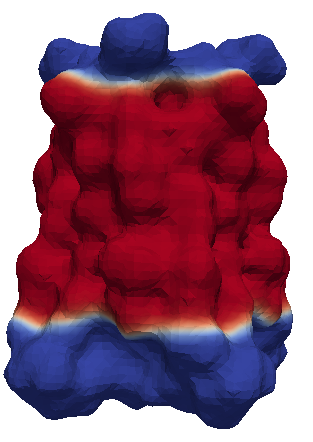}
\includegraphics[height=3.5cm]{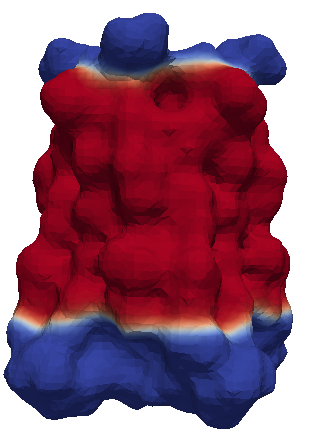}
\includegraphics[height=3.5cm]{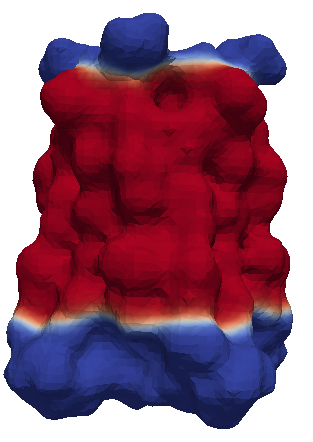}
\includegraphics[height=3.5cm]{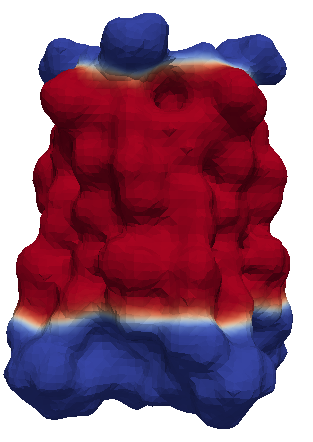} \\
\includegraphics[height=3.5cm]{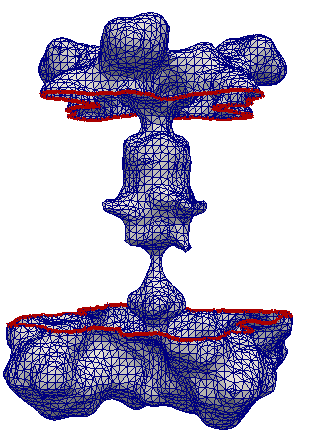} 
\includegraphics[height=3.5cm]{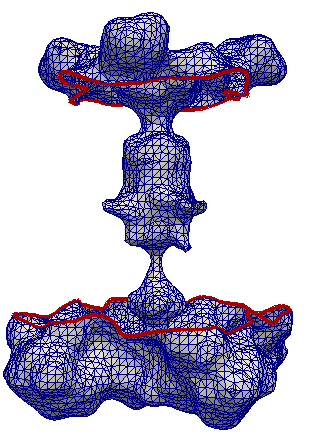} 
\includegraphics[height=3.5cm]{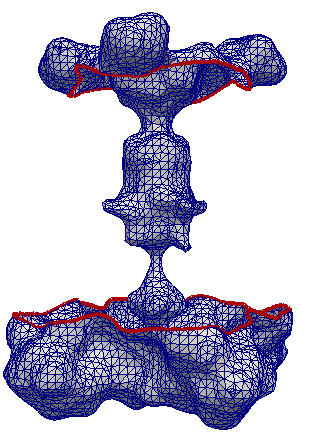} 
\includegraphics[height=3.5cm]{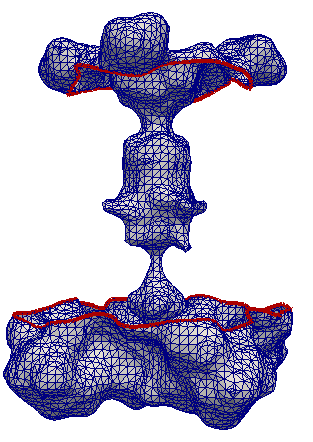} 
\includegraphics[height=3.5cm]{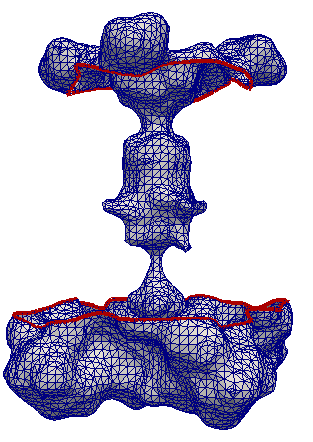} 
\caption{Snapshots in the surface phase boundary evolution on the proton channel protein (PDB ID: 2kqt) 
at time $t=0,20,40,60,80$. Initial phase boundaries are at $z=-10,13$. Surface triangular mesh is quasi-uniform with $12470$ vertices
and $24940$ triangles. Time increment $\Delta t=0.01$, $\xi=0.5$. Left: Plots of surface phase field function. 
Right: Surface mesh for the phase $\psi=-1$ highlights the evolution of the phase boundaries. 
The pole in the center represents the open proton channel.}
\label{fig:2kqt_case1}
\end{center}
\end{figure}

\begin{figure}[!ht]
\begin{center}
\includegraphics[height=3.5cm]{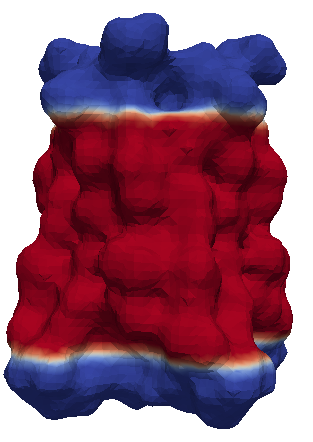}
\includegraphics[height=3.5cm]{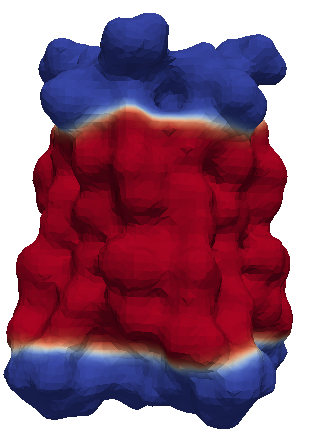}
\includegraphics[height=3.5cm]{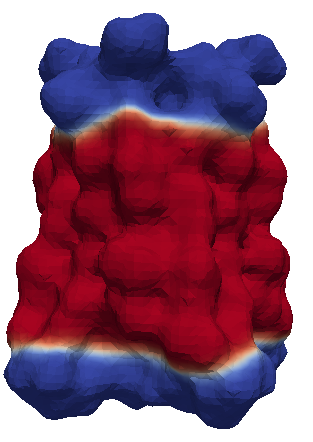}
\includegraphics[height=3.5cm]{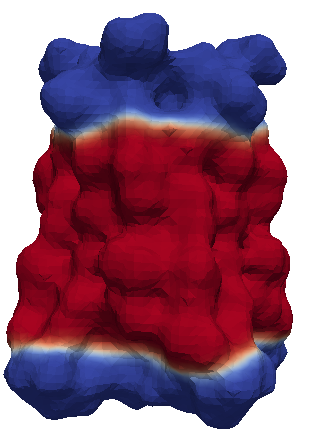}
\includegraphics[height=3.5cm]{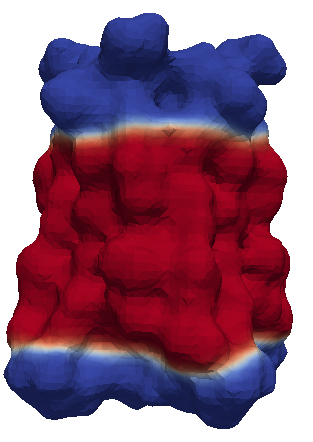} \\
\includegraphics[height=3.5cm]{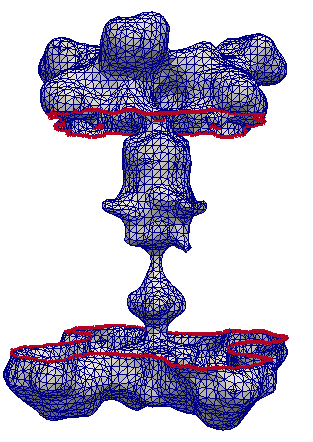} 
\includegraphics[height=3.5cm]{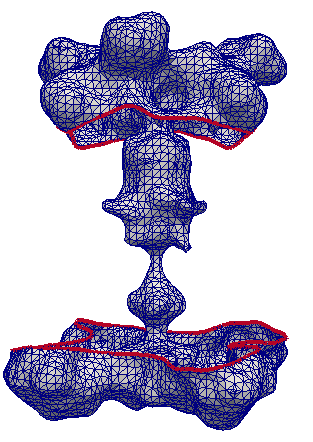}
\includegraphics[height=3.5cm]{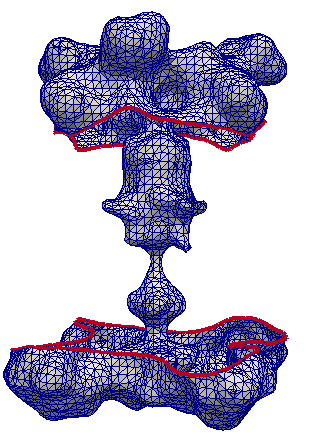} 
\includegraphics[height=3.5cm]{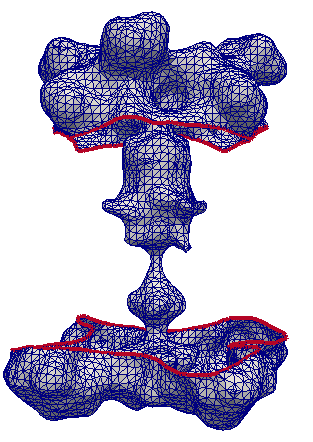}  
\includegraphics[height=3.5cm]{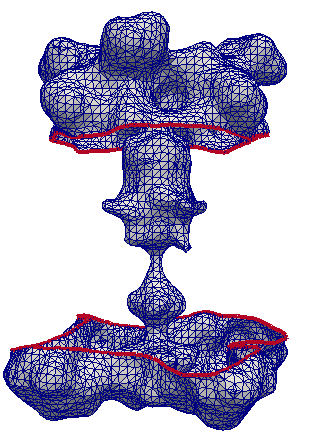} 
\caption{Snapshots in the surface phase boundary evolution on the proton channel protein (PDB ID: 2kqt) 
at time $t=0,20,40,60,80$. Initial phase boundaries are at $z=-13,11$. Mesh and parameters are identical to the case in 
Fig.~\ref{fig:2kqt_case1}. Left: Plots of surface phase field function. Right: Surface mesh for the phase $\psi=-1$ 
highlights the evolution of the phase boundaries.}
\label{fig:2kqt_case2}
\end{center}
\end{figure}

The third and fourth tests are carried out on the proton channel M2 (PDB ID: 2kqt). With a total of $3247$ atoms this protein 
has a rich surface morphology with many local minima of total geodesic curvature. Fig.~\ref{fig:2kqt_case1} shows the 
evolution of the two phase boundaries initially located at $z=-10,13$. The upper phase boundary moves up considerably to the local
minimum, while the lower phase boundary finds the local minimum geodesic energy after a slight adjustment of position. When the initial
phase boundaries are changed to be at $z=-13,11$, shown in Fig.~\ref{fig:2kqt_case2}, final phase boundaries different from the third
test are observed.

These four numerical examples illustrate that our model and numerical methods are very effective in capturing the initial 
condition-dependent local minimum of the geodesic curvature energy. In the applications to be 
presented in (\ref{sect:numerical_promemb}), we will scan over the full height of the membrane protein to 
set a range of initial phase boundaries and to locate the global minimum from the set of local minima.

\subsection{Validation of mappings to the middle plane}
\begin{figure}[!ht]
\begin{center}
\includegraphics[height=4.5cm]{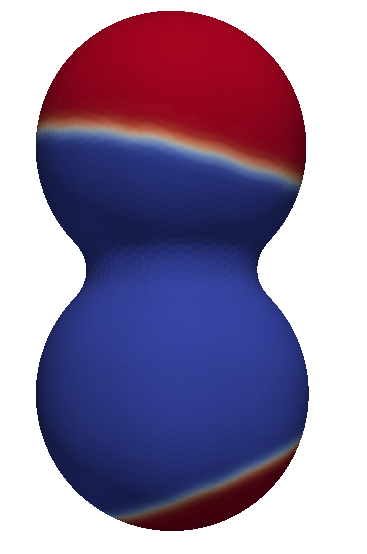}
\includegraphics[height=4.5cm]{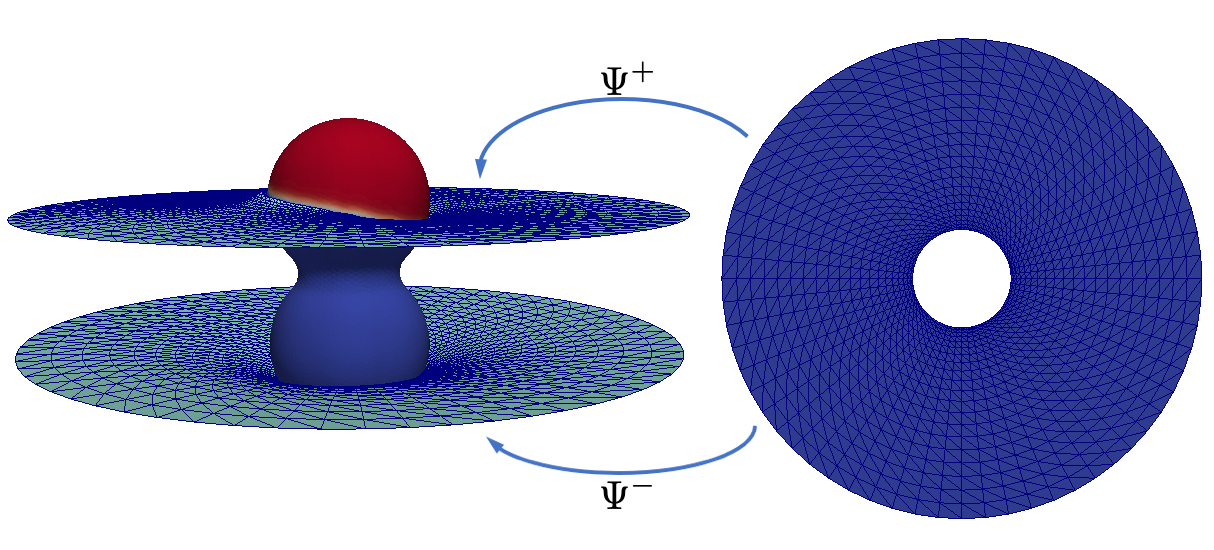}
\caption{Illustration of membrane surfaces mapped to the same middle annulus. Left: A model protein with two surface phase boundaries
representing the two contact curves with membrane surfaces. Right: The model protein is embedded in the bilayer represented as two membrane surfaces, whose
Monge parameterizations are defined on the planes mapped from the same annulus $\calR$ via respective transformations $\Psi^{\pm}$.}
\label{fig:2atm}
\end{center}
\end{figure}

We consider a model transmembrane protein and two phase boundaries on its surface modeling the protein-membrane interfaces, c.f.
Fig.~\ref{fig:2atm} (left). The common middle annulus is defined as follows. First, the geometrical centers and radii 
of the upper and lower phase boundaries are found. These define the position of the base planes for the two membrane surfaces. 
The average radii and average height ($z$-coordinate) will be used as the inner radius of middle annulus and its $z$-coordinate. 
This middle annulus is then discretized on the polar coordinate, with its uniformly discretized inner circle mapped to 
quasi-uniformly discretized upper and lower phase boundaries. Together with the mapping between the discretized exterior circles, 
we have the boundaries of the two base planes mapped to the middle annulus. 

Inhomogeneous Thompson-Thames-Mastin elliptic grid generator (\cite{Girdgeneration_KS}, Page 98) is adopted to 
produce interior mesh grids for the two base planes that are dense near the protein-membrane interfaces, 
c.f. Fig.~\ref{fig:2atm} (right). The quadrilateral grid meshes are then triangulated for the solution of 
Eq.(\ref{eqn:u-},\ref{eqn:u+}) using the interior penalty discontinuous Galerkin method described above. The 
vertices of the triangulated meshes for the two membrane surfaces inherit the one-one correspondence established 
on the mesh grids, allowing us to compute the coupling terms $(u^- - u^+)$ and $\Delta (u^- - u^+)$ 
in the two surface deformation equations. It is worth noting that by using the numerical mappings we do not change the 
physical domains or coordinate systems of the two membrane surfaces. The mappings are introduced merely to establish 
a one-one correspondence between the two membrane surfaces. Therefore one does not need to transform 
the surface deformation equations because of these numerical mappings of the computational domains.

\subsection{Membrane morphology induced by protein inclusion} \label{sect:numerical_promemb}
After validating the evolution of surface phase field function and the mapping to the middle annulus, we are now at
a position to apply these methodologies on the determination of membrane morphology induced by protein inclusion.
We consider a single M2 protein embedded in a homogeneous phosphatidylcholine (POPC) bilayer with equilibrium thickness $L_0=26\ang$. 
Although the height of protein is comparable to the bilayer thickness we will still expect a considerable compression of the bilayer 
in the vicinity of protein inclusion because the protein-membrane interfaces are highly curved. 
The mechanics and dielectric parameters of the POPC bilayer are taken from \cite{ArgudoD2017a}.
The outer radius of the middle common annulus is $100\ang$. The inner radius is about $12$ to $15 \ang$, 
depending on the position of the protein-membrane interfaces. The values of other parameters are $K_c=10.8, K_G =-9.8, 
\alpha = 0.004, K_{\alpha} = 0.33, k_t = \alpha$, all in the unit of $kcal/mol$.

\begin{figure}[!ht]
\begin{center}
\includegraphics[height=3.5cm]{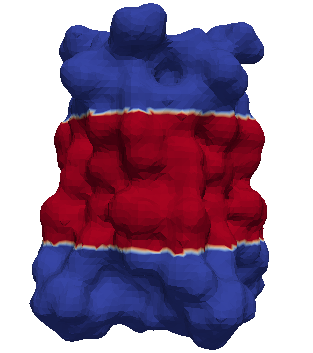}
\includegraphics[height=3.5cm]{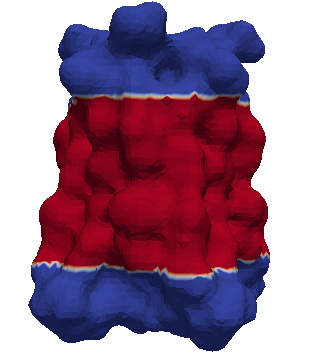}
\includegraphics[height=3.5cm]{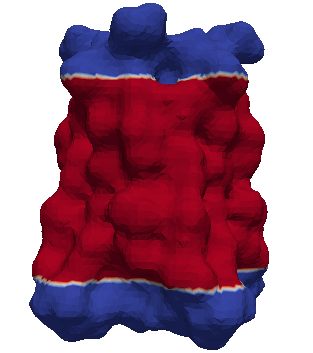}
\includegraphics[height=3.5cm]{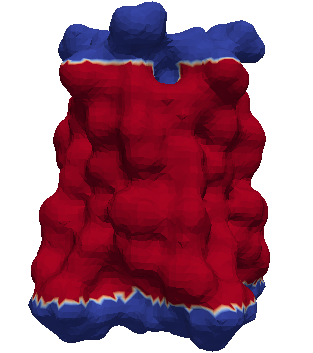} \\
\includegraphics[height=3.5cm]{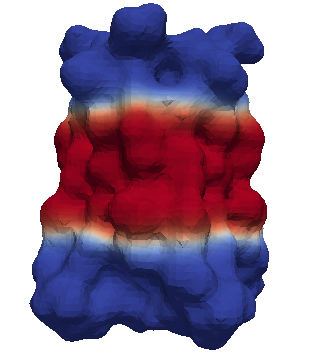}
\includegraphics[height=3.5cm]{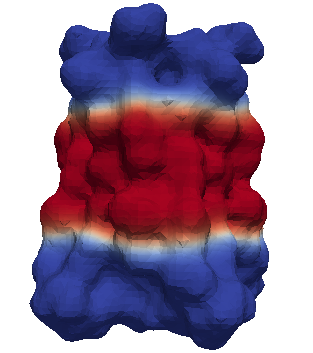} 
\includegraphics[height=3.5cm]{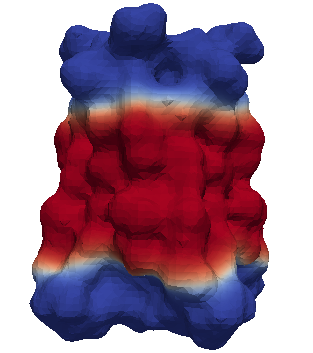}
\includegraphics[height=3.5cm]{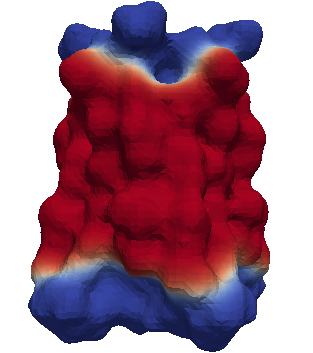}
\caption{Surface phase fields with different initial phase boundaries (top) and the corresponding equilibrium fields (bottom).
From left to right: $z_{ini}=\pm8, \pm10, \pm 12, \pm 14 \ang$.}
\label{fig:2kqt_surfpot}
\end{center}
\end{figure}

\begin{figure}[!ht]
\begin{center}
\includegraphics[height=3.3cm]{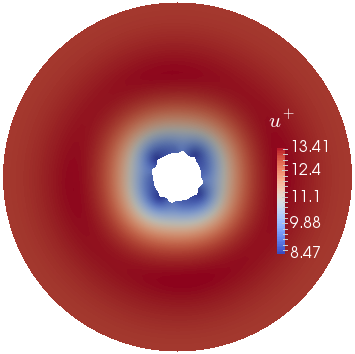}
\includegraphics[height=3.3cm]{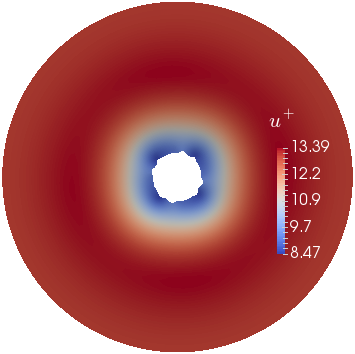}
\includegraphics[height=3.3cm]{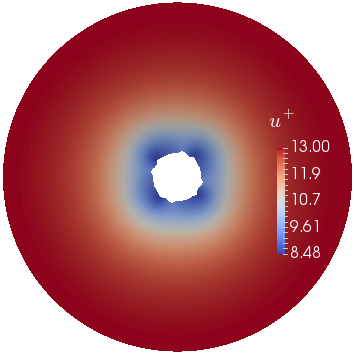}
\includegraphics[height=3.3cm]{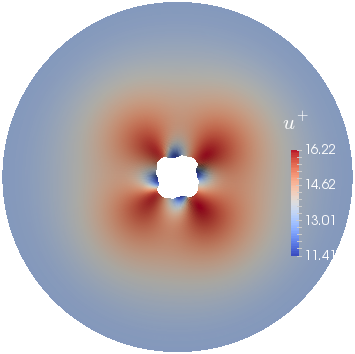} \\
\includegraphics[height=3.3cm]{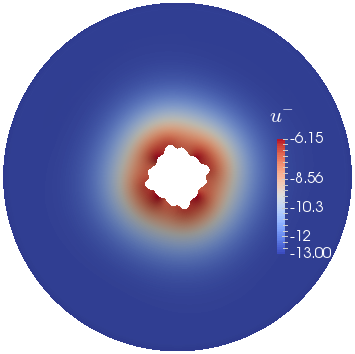}
\includegraphics[height=3.3cm]{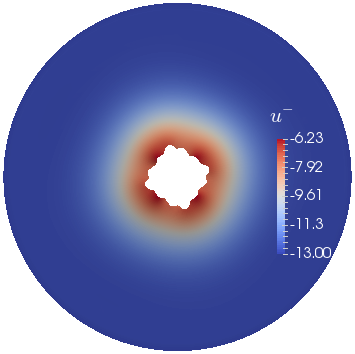} 
\includegraphics[height=3.3cm]{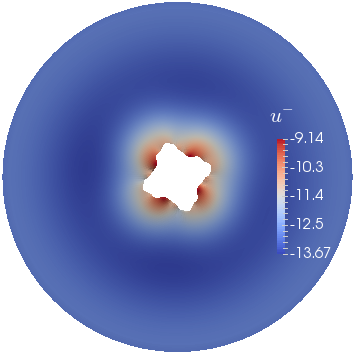}
\includegraphics[height=3.3cm]{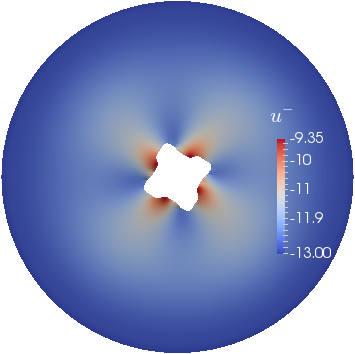}
\caption{Equilibrium top ($u^+)$ and bottom ($u^-$) membrane surfaces obtained with different initial phase field boundaries at
$z=\pm8, \pm10, \pm 12, \pm 14 \ang$, from left to right.}
\label{fig:2kqt_memb_deform}
\end{center}
\end{figure}

We first look at the protein-membrane interfaces modelled as the phase boundaries with different surface phase field 
initializations $z_{ini}=\pm 8 \sim \pm 14 \ang$. As shown in Fig.~\ref{fig:2kqt_surfpot},
the top phase boundary does not change when the initial position is placed at $8 \le z_{ini} \le 12.5 \ang$ as they will be trapped 
to the same position with a local minimum geodesic curvature energy is observed. The bottom interface stays at the same local 
minimum for $-11.3 \le z_{ini} \le -8 \ang$, and then moves to another local minimum for a lower initial phase boundary. We will 
consider these initial protein-membrane 
interfaces and apply Algorithm \ref{alg:full_alg} to find the equilibrium membrane surfaces with minimum total energy, and 
the results are shown in Fig.~\ref{fig:2kqt_memb_deform}.
For $z_{ini} =\pm 8 \ang$ or $\pm 10 \ang$, the corresponding equilibrated protein-membrane interfaces are identical and 
located near $z=-6 \ang$ and $z=8 \ang$, respectively. The two membrane surfaces have to bend significantly to match these 
inner boundaries as their exterior boundaries are clamped at $z=\pm 13 \ang$. This shall generate large bending and 
compression energies, see also Table \ref{tab:energies}. When the initial phase boundaries are placed further 
apart at $|z_{ini}| >11.3 \ang$, the new bottom protein-membrane interface promotes the curving of the membrane 
near the protein, shown as the peaking of saddle splay energy in the table, see also the variation of the membrane
heights near the protein in Fig.~\ref{fig:2kqt_memb_deform}, where the four peaks of the displacement signifies
the quatermer structure of M2.
\begin{table}[!ht]
\begin{center}
\begin{tabular}{rlllllll} \hline
   $z_{ini}$     &  $\pm 8$  & $\pm 9$ & $\pm 10$ & $\pm 11$ & $\pm 12$ &  $\pm  13$ &  $\pm 14$ \\ \hline
Splay             & -310.7   & -305.6   & -292.1    & -350.2     &  -365.6     & 45.72     & -183.23  \\
Saddle splay      &  2.11    & 1.13    & 0.96     & -6.87     & -13.56     & 27.54       & 6.99 \\
Surface tension   &  1.50     & 1.02    & 0.65     & 0.36     & 0.24     & 0.60       & 0.54 \\
Compression       & 196159   & 136287    & 88400     & 47046    & 16106     & 534.2   &  8382 \\
Tilt-stretch      & 0.75    & 0.51    &  0.32       &  0.19    & 0.13     &  0.20      & 0.24 \\ \hline
\end{tabular}
\caption{Energies in $kcal/mol$ corresponding to different modes of bilayer deformation when the initial membrane-protein contact 
curves are varied.}
\label{tab:energies}
\end{center}
\end{table}

\begin{figure}[!ht]
\begin{center}
\includegraphics[height=2.5cm]{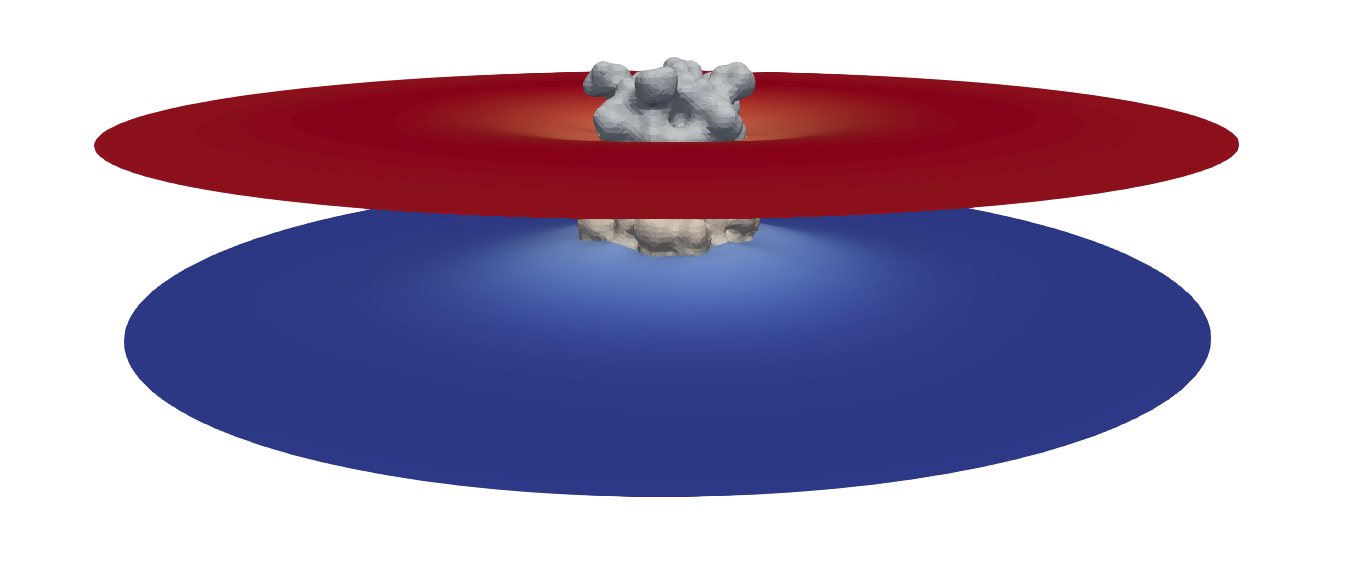}
\includegraphics[height=2.5cm]{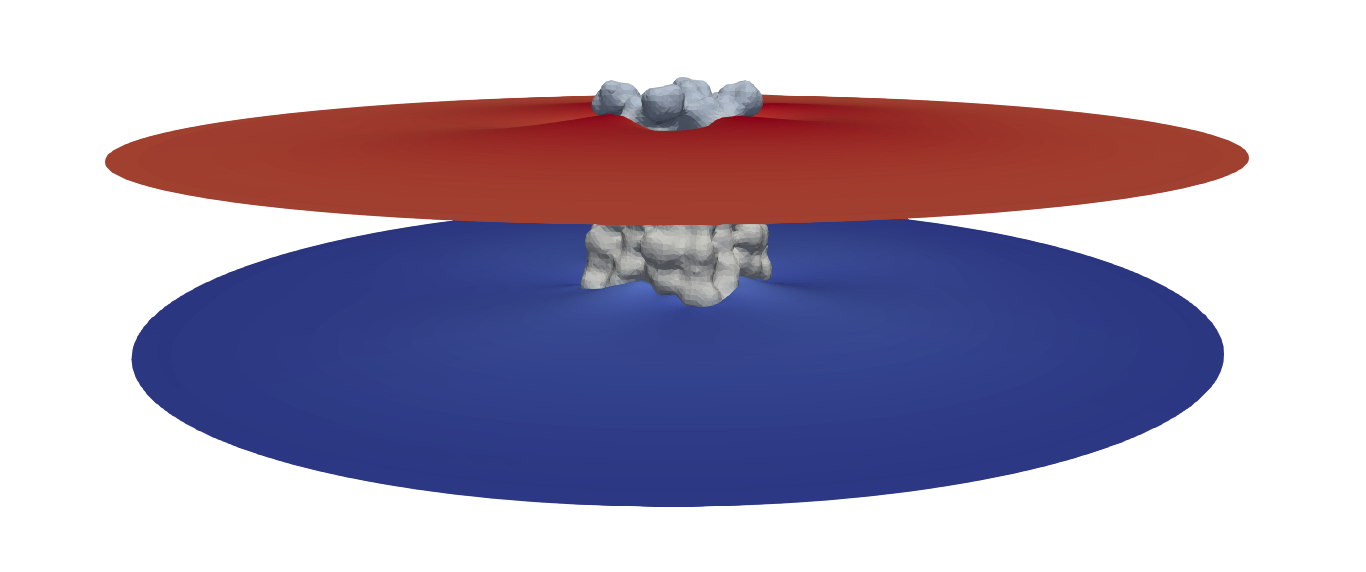}
\caption{Embedding of M2 channel in the bilayer with $z_{ini} =\pm 8$ (left) and $z_{ini}=\pm 13$ (right).}
\label{fig:complex_deform}
\end{center}
\end{figure}

The significant increase of bilayer compression near the embedded protein from $z_{ini}=\pm 8 \ang$ to $z_{ini}=\pm 13 \ang$ 
is also illustrated in Fig.~\ref{fig:complex_deform}. Maximum energy for saddle splay is observed at $z_{ini}=\pm 13 \ang$,
indicating the largest negative Gaussian curvature is generated, a feature that is well observed in small angle X-ray scattering
and molecular dynamics simulations \cite{SchmidtN2013a,PaulinoJ2019a}. However, this negative 
Gaussian curvature is confined to a small neighborhood near the embedded protein. 
We attribute this confinement 
to the homogeneous Dirichlet (aka clamped) boundary condition (\ref{eqn:BC_far}). In experiments the far boundary is not 
clamped while in molecular dynamics simulations a periodic boundary condition is usually chosen. A more realistic boundary 
condition on the exterior edge of the annulus would be
\begin{equation}
\Delta u = 0, \quad \frac{\partial \Delta u}{\partial n} = 0,
\end{equation}
i.e., the membrane is stress-free there. However, there does not exist a robust numerical method for biharmonic 
equations with mixed boundary conditions in arbitrarily complicated 2-D domain. Recently emerged weak Galerkin 
method might posses the flexibility of approximating forth-order equations with mixed boundary conditions 
(Dirichlet boundary conditions (\ref{eqn:BC_near}) on the inner edge and stress-free boundary condition 
on the exterior edge).

\section{Conclusion} \label{sect:summary}
In this paper we propose a geodesic curvature energy model for characterizing the protein-membrane interfaces. These
interfaces represent the essential information of the boundaries in the continuum or hybrid modeling of bilayer membrane 
morphology induced by embedded proteins. Along with a surface phase field approximation of the geodesic curvature, our 
efforts present a computationally tractable geometrical characterization of the boundary conditions for the protein-membrane
interactions. To further completely couple the surface morphology of two leaflets we first envision an annulus located between two
surfaces and then construct conformal mapping between individual surface and the middle annulus through numerical grid generation.
We integrate these two new features into a general energetic functional model of protein-membrane interactions. Numerical experiments
demonstrate that our methods are efficient and robust in locating the highly complicated protein-membrane interfaces and the 
corresponding membrane morphology. 

Our work can be improved and extended mathematically and computationally 
in several different directions. A stress-free boundary condition on the exterior edge of membrane shall better reflect the local mechanical 
constraint and we are currently developing a weak Galerkin finite element method for solving the resulting fourth order equations with mixed boundary conditions. 
Secondly, the manual scanning on the protein surface for locating a global minimum of total geodesic curvature can be replaced by a surface 
Allan-Cahn equation with stochastic forcing term (or a stochastic term directly added to the total geodesic curvature energy) that could provide an 
additional force driving the evolving surface phase boundary from a local minimum to the global minimum. Finally, force and torque balance equations
can be introduced to the transmembrane protein to better describe the tilting and orientation of the embedded protein.

\section*{Acknowledgements}
This work has been partially supported by National Institutes of Health through the grant R01GM117593 as 
part of the joint DMS/NIGMS initiative to support research at the interface of the biological and mathematical sciences.



\end{document}